\documentclass[thmsa]{article}
\usepackage{sw20lart}

\input{tcilatex}
\begin{document}

\author{S. Manoff \\
%EndAName
\textit{Bulgarian Academy of Sciences}\\
\textit{Institute for Nuclear Research and Nuclear Energy}\\
\textit{Department of Theoretical Physics}\\
\textit{Blvd. Tzarigradsko Chaussee 72}\\
\textit{1784 Sofia - Bulgaria} \and R. Lazov \\
%EndAName
\textit{Bulgarian Academy of Sciences}\\
\textit{Institute for Mathematics and Informatics}\\
\textit{G. Bonchev Str. 7}\\
\textit{1300 Sofia - Bulgaria}}
\title{\textsc{Projections and covariant divergency}\\
\textsc{of the energy-momentum tensors}}
\date{\textit{e-mail address}: smanov@inrne.bas.bg }
\maketitle

\begin{abstract}
The invariant projections of the energy-momentum tensors of Lagrangian
densities for tensor fields over differentiable manifolds with contravariant
and covariant affine connections and metrics [$(\overline{L}_n,g)$-spaces]
are found by the use of an non-null (non-isotropic) contravariant vector
field and its corresponding projective metrics. The notions of rest mass
density, momentum density, energy current density and stress tensor are
introduced as generalizations of these notions from the relativistic
continuum media mechanics. The energy-momentum tensors are represented by
means of the introduced notions and the corresponding identities are found.
The notion of covariant differential operator along a contravariant tensor
field is introduced. On its basis, as a special case, the notion of
contravariant metric differential operator is proposed. The properties of
the operators are considered. By the use of these operators the notion of
covariant divergency of a mixed tensor field is determined. The covariant
divergency of tensor fields of second rank of the types 1 and 2 is found.
Invariant representations of the covariant divergency of the energy-momentum
tensor are obtained by means of the projective metrics of a contravariant
non-isotropic (non-null) vector field and the corresponding rest mass
density, momentum density, and energy flux density. An invariant
representation of the first Noether identity is found as well as relations
between the covariant divergencies of the different energy-momentum tensors
and their structures determining covariant local conserved quantities.
\end{abstract}

\section{Invariant projections of a mixed tensor field of second rank}

In the relativistic continuum media mechanics notions are introduced as
generalizations of the same notions of the classical continuum media
mechanics. This has been done by means of the projections of the (canonical,
symmetric of Belinfante or symmetric of Hilbert) energy-momentum tensors
along or orthogonal to a non-isotropic (non-null) contravariant vector field.

There are possibilities for using the projections for finding out the
physical interpretations of the determined energy-momentum tensors. In an
analogous way as in (pseudo) Riemannian spaces without torsion ($V_n$%
-spaces), the different relations between the quantities with well known
physical interpretations can be considered as well as their application for
physical systems described by means of mathematical models over
differentiable manifold with affine connections and metrics [$(\overline{L}%
_n,g)$-spaces].

The energy-momentum tensors are obtained as mixed tensor fields of second
rank of type 1 [by the use of the procedure on the grounds of the method of
Lagrangians with covariant derivatives (MLCD) \cite{Manoff-0}] 
\begin{equation}  \label{IX.1.-1}
G=G_\alpha \,^\beta \cdot e_\beta \otimes e^\alpha =G_i\,^j\cdot \partial
_j\otimes dx^i
\end{equation}

\noindent in contrast to the mixed tensor fields of second rank of type 2 
\begin{equation}  \label{IX.1.-2}
\overline{G}=\overline{G}\,^\beta \,_\alpha \cdot e^\alpha \otimes e_\beta =%
\overline{G}\,^j\,_i\cdot dx^i\otimes \partial _j\text{ .}
\end{equation}

$\{e^\alpha \}$ and $\{e_\alpha \}$ are non-co-ordinate (non-holonomic)
covariant and contravariant basic vector fields respectively, $(\alpha
,\beta =1,...,n)$, $\{dx^i\}$ and $\{\partial _i\}$ are co-ordinate
(holonomic) covariant and contravariant basic vector fields respectively $%
(i,j=1,...,n)$, $\dim (\overline{L}_n,g)=n$.

To every covariant basic vector field a contravariant basic vector field can
be juxtaposed and vice versa by the use of the contravariant and covariant
metric tensor fields 
\begin{equation}  \label{IX.1.-3}
g(e_\gamma )=g_{\alpha \overline{\gamma }}\cdot e^\alpha \text{ ,\thinspace
\thinspace \thinspace \thinspace \thinspace \thinspace }g(\partial _j)=g_{i%
\overline{j}}\cdot dx^i\text{ ,\thinspace \thinspace \thinspace \thinspace }%
\overline{g}(e^\gamma )=g^{\alpha \overline{\gamma }}\cdot e_\alpha \text{
,\thinspace \thinspace \thinspace \thinspace }\overline{g}(dx^j)=g^{i%
\overline{j}}\cdot \partial _i\text{ .}
\end{equation}

On this basis a tensor field of the type 2 can be related to a tensor field
of the type 1 by the use of the covariant and contravariant tensor fields $%
g=g_{ij}\cdot dx^i.dx^j$ and $\overline{g}=g^{ij}\cdot \partial _i.\partial
_j$ [$dx^i.dx^j=(1/2)(dx^i\otimes dx^j+dx^j\otimes dx^i)$, $\partial
_i.\partial _j=(1/2)(\partial _i\otimes \partial _j+\partial _j\otimes
\partial _i)$] 
\begin{equation}  \label{IX.1.-5}
\overline{G}=g(G)\overline{g}=\overline{G}\,^\beta \,_\alpha \cdot e^\alpha
\otimes e_\beta =g_{\alpha \overline{\gamma }}\cdot G_\delta \,^\gamma \cdot
g^{\overline{\delta }\beta }\cdot e^\alpha \otimes e_\beta \text{ ,}
\end{equation}

\begin{equation}  \label{IX.1.-6}
\overline{G}\,^\beta \,_\alpha =g_{\alpha \overline{\gamma }}\cdot G_\delta
\,^\gamma \cdot g^{\overline{\delta }\beta }\text{ ,}
\end{equation}

\begin{equation}  \label{IX.1.-7}
G=\overline{g}(\overline{G})g=G_\alpha \,^\beta \cdot e_\beta \otimes
e^\alpha =g^{\beta \overline{\delta }}\cdot \overline{G}\,^\gamma \,_\delta
\cdot g_{\overline{\gamma }\alpha }\cdot e_\beta \otimes e^\alpha \text{ ,}
\end{equation}

\begin{equation}  \label{IX.1.-8}
G_\alpha \,^\beta =g^{\beta \overline{\delta }}\cdot \overline{G}\,^\gamma
\,_\delta \cdot g_{\overline{\gamma }\alpha }\text{ .}
\end{equation}

The Kronecker tensor field appears as a mixed tensor field of second rank of
the type 1 
\[
Kr=g_\beta ^\alpha \cdot e_\alpha \otimes e^\beta =g_j^i\cdot \partial
_i\otimes dx^j\text{ } 
\]

\noindent and can be projected by means of the non-isotropic (non-null)
contravariant vector field $u$ and its projection metrics $h_u$ and $h^u$ [$%
h_u=g-\frac 1e\cdot g(u)\otimes g(u)$, $h^u=\overline{g}-\frac 1e\cdot
u\otimes u$, $e=g(u,u)\neq 0$] 
\[
Kr=\varepsilon _{Kr}\cdot u\otimes g(u)+u\otimes g(^{Kr}\pi
)+\,^{Kr}s\otimes g(u)+(^{Kr}S)g\text{ ,} 
\]

\noindent where 
\begin{equation}  \label{IX.1.-42}
\begin{array}{c}
\varepsilon _{Kr}=\frac 1{e^2}\cdot [g(u)](Kr)u=\frac 1{e^2}.u_{\overline{%
\alpha }}\cdot u^{\overline{\alpha }}=\frac 1{e^2}.g_{\overline{\alpha }%
\overline{\beta }}\cdot u^{\overline{\alpha }}\cdot u^\beta = \\ 
=\frac 1{e^2}.f^\rho \,_\beta \cdot f^\sigma \,\cdot f^\beta \,_\delta \cdot
g_{\rho \sigma }\cdot u^\gamma \cdot u^\delta =\frac 1{e^2}.u^{\overline{i}%
}\cdot u_{\overline{i}}=\frac 1{e^2}\cdot g_{\overline{i}\overline{k}}\cdot
u^{\overline{i}}\cdot u^k=\frac 1e\cdot k\text{ ,}
\end{array}
\end{equation}

\begin{equation}  \label{IX.1.-43}
k=\frac 1e\cdot [g(u)](Kr)u=\frac 1e.u^{\overline{i}}\cdot u_{\overline{i}}%
\text{ ,\thinspace \thinspace \thinspace \thinspace \thinspace }u_{\overline{%
i}}=g_{\overline{i}\overline{j}}\cdot u^j\text{ ,}
\end{equation}

\begin{equation}  \label{IX.1.-44}
\begin{array}{c}
^{Kr}\pi =\frac 1e\cdot [g(u)](Kr)h^u=\frac 1e\cdot g_{\overline{\beta }%
\overline{\gamma }}\cdot u^\gamma \cdot g_\delta ^\beta \cdot h^{\overline{%
\delta }\alpha }\cdot e_\alpha =\frac 1e\cdot g_{\overline{\beta }\overline{%
\gamma }}\cdot h^{\overline{\beta }\alpha }\cdot u^\gamma \cdot e_\alpha =
\\ 
=\frac 1e\cdot u_{\overline{\beta }}\cdot h^{\overline{\beta }\alpha }\cdot
e_\alpha =\,^{Kr}\pi ^\alpha \cdot e_\alpha \text{ ,}
\end{array}
\end{equation}

\begin{equation}  \label{IX.1.-45}
^{Kr}\pi ^\alpha =\frac 1e\cdot g_{\overline{\beta }\overline{\gamma }}\cdot
u^\gamma \cdot h^{\overline{\beta }\alpha }=\frac 1e\cdot u_{\overline{\beta 
}}\cdot h^{\overline{\beta }\alpha }\text{ ,\thinspace \thinspace \thinspace
\thinspace \thinspace \thinspace \thinspace }^{Kr}\pi ^i=\frac 1e\cdot g_{%
\overline{k}\overline{l}}\cdot u^l\cdot h^{\overline{k}i}=\frac 1e\cdot u_{%
\overline{k}}\cdot h^{\overline{k}i}\text{ ,\thinspace }
\end{equation}

\begin{equation}  \label{IX.1.-47}
^{Kr}s=\frac 1e\cdot h^u(g)(Kr)(u)=\,^{Kr}s^\alpha \cdot e_\alpha
=\,^{Kr}s^i\cdot \partial _i\text{ ,}
\end{equation}
\begin{equation}  \label{IX.1.-49}
^{Kr}s^\alpha =\frac 1e\cdot h^{\alpha \beta }\cdot g_{\overline{\beta }%
\overline{\delta }}\cdot u^{\overline{\delta }}\text{ ,\thinspace \thinspace
\thinspace \thinspace \thinspace \thinspace \thinspace \thinspace \thinspace 
}^{Kr}s^i=\frac 1e\cdot h^{ik}\cdot g_{\overline{k}\overline{l}}\cdot u^{%
\overline{l}}\text{ ,}
\end{equation}
\begin{equation}  \label{IX.1.-50}
^{Kr}S=h^u(g)(Kr)h^u=\,^{Kr}S^{\alpha \beta }\cdot e_\alpha \otimes e_\beta
=\,^{Kr}S^{ij}\cdot \partial _i\otimes \partial _j\text{ ,}
\end{equation}
\begin{equation}  \label{IX.1.-51}
^{Kr}S^{\alpha \beta }=h^{\alpha \gamma }\cdot g_{\overline{\gamma }%
\overline{\delta }}\cdot h^{\overline{\delta }\beta }\text{ ,\thinspace
\thinspace \thinspace \thinspace \thinspace \thinspace }^{Kr}S^{ij}=h^{ik}%
\cdot g_{\overline{k}\overline{l}}\cdot h^{\overline{l}j}\text{ ,}
\end{equation}
\[
h^u[g(u)]=[g(u)](h^u)=0\text{ ,\thinspace \thinspace \thinspace \thinspace
\thinspace \thinspace \thinspace }h^u(g)h^u=h^u\text{ ,} 
\]
\begin{equation}  \label{IX.1.-52}
\begin{array}{c}
(Kr) \overline{g}=\frac ke\cdot u\otimes u+u\otimes \,^{Kr}\pi
+\,^{Kr}s\otimes u+\,^{Kr}S\text{ ,} \\ 
(Kr)\overline{g}=g^{\overline{\alpha }\beta }\cdot e_\alpha \otimes e_\beta
=g^{\overline{i}j}\cdot \partial _i\otimes \partial _j\text{ .}
\end{array}
\end{equation}

The corresponding to the Kronecker tensor field mixed tensor field of the
type 2 
\begin{equation}  \label{IX.1.-53}
\overline{K}r=g(Kr)\overline{g}=g_{\alpha \overline{\gamma }}\cdot g^{%
\overline{\gamma }\beta }\cdot e^\alpha \otimes e_\beta =g_{i\overline{k}%
}\cdot g^{\overline{k}j}\cdot dx^i\otimes \partial _j\text{ }
\end{equation}

\noindent does not appear in general as a Kronecker tensor field.

\textit{Special case}: $S=C:f^i\,_j=g_j^i$, \thinspace ($f^\alpha \,_\beta
=g_\beta ^\alpha $): 
\begin{equation}  \label{IX.1.54-59}
\begin{array}{c}
\overline{K}r=g_\beta ^\alpha \cdot e^\beta \otimes e_\alpha =g_j^i\cdot
dx^i\otimes \partial _j\text{ ,\thinspace \thinspace \thinspace } \\ 
k=1 \text{,\thinspace \thinspace \thinspace \thinspace \thinspace \thinspace 
}\varepsilon _{Kr}=\frac 1e\text{,\thinspace \thinspace \thinspace
\thinspace }^{Kr}\pi =0\text{ ,\thinspace \thinspace \thinspace \thinspace
\thinspace \thinspace }^{Kr}s=0\text{ ,\thinspace \thinspace \thinspace
\thinspace \thinspace }^{Kr}S=h^u\text{ ,} \\ 
Kr=\frac 1e\cdot u\otimes g(u)+(h^u)g=\overline{g}(g)\text{ ,\thinspace
\thinspace \thinspace \thinspace \thinspace \thinspace }(Kr)\overline{g}=%
\overline{g}\text{ .}
\end{array}
\end{equation}

The representation of the tensor fields of the type 1 by the use of the
non-isotropic (non-null) contravariant vector field $u$ and its projective
metrics $h^u$ and $h_u$ corresponds in its form to the representation of the
viscosity tensor and the energy-momentum tensors in the continuum media
mechanics in $V_3$- or $V_4$-spaces, where $\varepsilon _G$ is the inner
energy density, $^G\pi $ is the conductive momentum, $e.^Gs$ is the
conductive energy flux density and $^GS$ is the stress tensor density. An
analogous interpretation can also be accepted for the projections of the
energy-momentum tensors found by means of the method of Lagrangians with
covariant derivatives (MLCD).

\section{Energy-momentum tensors and the rest mass density}

The covariant Noether identities (generalized covariant Bianchi identities)
can be considered as identities for the components of mixed tensor fields of
second rank of the first type. The second covariant Noether identity 
\[
\overline{\theta }_\alpha \,^\beta -\,_sT_\alpha \,^\beta \equiv \overline{Q}%
\,_\alpha \,^\beta 
\]

\noindent can be written in the form 
\begin{equation}  \label{IX.2.-1}
\theta -\,_sT\equiv Q\text{ ,}
\end{equation}

\noindent where 
\begin{equation}  \label{IX.2.-2}
\begin{array}{c}
\theta = \overline{\theta }_\alpha \,^\beta \cdot e_\beta \otimes e^\alpha =%
\overline{\theta }_i\,^j\cdot \partial _j\otimes dx^i\text{ ,} \\ 
_sT=\,_sT_\alpha \,^\beta \cdot e_\beta \otimes e^\alpha =\,_sT_i\,^j\cdot
\partial _j\otimes dx^i \text{ ,} \\ 
Q=\overline{Q}_\alpha \,^\beta \cdot e_\beta \otimes e^\alpha =\overline{Q}%
_i\,^j\cdot \partial _j\otimes dx^i\text{ ,}
\end{array}
\end{equation}

$\theta $ is the generalized canonical energy-momentum tensor (GC-EMT) of
the type 1; $_sT$ is the symmetric energy-momentum tensor of Belinfante
(S-EMT-B) of the type 1; $Q$ is the variational energy-momentum tensor of
Euler-Lagrange (V-EMT-EL) of the type 1.

The second covariant Noether identity for the energy-momentum tensors of the
type 1 is called \textit{second covariant Noether identity of type 1.}

By means of the non-isotropic contravariant vector field $u$ and its
corresponding projective metric the energy-momentum tensors can be
represented in an analogous way as the mixed tensor fields of the type 1.

The structure of the generalized canonical energy-momentum tensor and the
symmetric energy-momentum tensor of Belinfante for the metric and non-metric
tensor fields has similar elements and they can be written in the form 
\begin{equation}  \label{IX.2.-3}
\begin{array}{c}
G=\,_kG-L\cdot Kr \text{ ,} \\ 
\theta =\,_k\theta -L\cdot Kr\text{ ,\thinspace \thinspace \thinspace
\thinspace \thinspace \thinspace \thinspace \thinspace \thinspace \thinspace
\thinspace }_sT=\mathit{T}-L\cdot Kr\text{ ,}
\end{array}
\end{equation}

\noindent where 
\begin{equation}  \label{IX.2.-6}
_k\theta =\,_k\overline{\theta }_\alpha \,^\beta \cdot e_\beta \otimes
e^\alpha =\,_k\overline{\theta }_i\,^j\cdot \partial _j\otimes dx^i\text{ .}
\end{equation}

On the analogy of the notions of the continuum media mechanics $_kG$ is
called \textit{viscosity tensor field}.

$G$ and $_kG$ can be written by means of $u$, $h^u$ and $h_u$ in the form 
\begin{equation}  \label{IX.2.-9}
\begin{array}{c}
G=\varepsilon _G\cdot u\otimes g(u)+u\otimes g(^G\pi )+\,^Gs\otimes
g(u)+\,(^GS)g \text{ ,} \\ 
_kG=\varepsilon _k\cdot u\otimes g(u)+u\otimes g(^k\pi )+\,^ks\otimes
g(u)+\,(^kS)g\text{ .}
\end{array}
\end{equation}

From the relation (\ref{IX.2.-3}) the relations between the different
projections of $G$ and $_kG$ follow. If we introduce the abbreviation $%
\varepsilon _G=\rho _G$, then 
\begin{equation}  \label{IX.2.-10}
\begin{array}{c}
\varepsilon _k=\rho _G+\frac 1e\cdot L\cdot k \text{ , \thinspace \thinspace
\thinspace }^k\pi =\,^G\pi +L\cdot ^{Kr}\pi \text{ ,\thinspace \thinspace
\thinspace \thinspace \thinspace }^ks=\,^Gs+L\cdot ^{Kr}s\text{ ,} \\ 
^kS=\,^GS+L\cdot ^{Kr}S\text{ ,}
\end{array}
\end{equation}

\noindent and $G$ can be written by means of (\ref{IX.2.-3}) in the form 
\begin{equation}  \label{IX.2.-14}
G=(\rho _G+\frac 1e\cdot L\cdot k)\cdot u\otimes g(u)-L\cdot Kr+u\otimes
g(^k\pi )+\,^ks\otimes g(u)+\,(^kS)g\text{ ,}
\end{equation}

\noindent where 
\[
\rho _G=\frac 1{e^2}\cdot [g(u)](G)(u) 
\]

\noindent is the \textit{rest mass density }of the energy-momentum tensor $G$
of the type 1. This type of representation of a given energy-momentum tensor 
$G$ by means of the projective metrics of $u$ and $\rho _G$ is called 
\textit{representation of }$G$\textit{\ by means of the projective metrics
of the contravariant non-isotropic (non-null) vector field }$u$\textit{\ and
the rest mass density} $\rho _G$.

There are other possibilities for representation of $G$ by means of $u$ and
its corresponding projective metrics.

If we introduce the abbreviation 
\begin{equation}  \label{IX.2.-15}
p_G=\rho _G\cdot u+\,^G\pi \text{ ,}
\end{equation}

\noindent where $p_G$ is the \textit{momentum density} of the
energy-momentum tensor $G$ of the type 1, then $G$ can be written in the
form 
\begin{equation}  \label{IX.2.-16}
\begin{array}{c}
G=u\otimes g(\rho _G\cdot u+\,^G\pi )+\,^Gs\otimes g(u)+(^GS)g \text{ ,} \\ 
G=u\otimes g(p_G)+\,^Gs\otimes g(u)+(^GS)g\text{ .}
\end{array}
\end{equation}

The representation of $G$ by means of the last relation is called \textit{%
representation of }$G$\textit{\ by means of the projective metrics of the
contravariant non-isotropic contravariant vector field }$u$\textit{\ and the
momentum density} $p_G$.

By the use of the relations 
\begin{equation}  \label{IX.2.-17}
g(^G\pi ,u)=0\text{ ,\thinspace \thinspace \thinspace \thinspace \thinspace
\thinspace \thinspace \thinspace \thinspace }(^GS)[g(u)]=0\text{ ,}
\end{equation}

\noindent valid (because of their constructions) for every energy-momentum
tensor $G$ and the definitions 
\begin{equation}  \label{IX.2.-18}
e_G=G(u)=(G)(u)=e\cdot (\rho _G\cdot u+\,^Gs)\text{ ,\thinspace \thinspace
\thinspace \thinspace }g(u,u)=e\neq 0\text{ ,}
\end{equation}
where $e_G$ is the \textit{energy flux density} of the energy-momentum
tensor $G $ of the type 1, the tensor field $G$ can be written in the form 
\begin{equation}  \label{IX.2.-19}
\begin{array}{c}
G=(\rho _G\cdot u+\,^Gs)\otimes g(u)+u\otimes g(^G\pi )+(^GS)g \text{ ,} \\ 
G=\frac 1e\cdot e_G\otimes g(u)+u\otimes g(^G\pi )+(^GS)g\text{ .}
\end{array}
\end{equation}

The representation of $G$ by means of the last expression is called \textit{%
representation of }$G$\textit{\ by means of the projective metrics of the
contravariant non-isotropic vector field }$u$\textit{\ and the energy flux
density} $e_G$.

The generalized canonical energy-momentum tensor $\theta $ can be
represented, in accordance with the above described procedure, by the use of
the projective metrics of $u$ and the rest mass density $\rho _\theta $%
\begin{equation}  \label{IX.2.-20}
\theta =\,_k\theta -L\cdot Kr\text{ ,\thinspace \thinspace \thinspace
\thinspace \thinspace \thinspace \thinspace }_k\theta =\theta +L\cdot Kr%
\text{ ,}
\end{equation}
\begin{equation}  \label{IX.2.-21}
\theta =(\rho _\theta +\frac 1e\cdot L\cdot k)\cdot u\otimes g(u)-L\cdot
Kr+u\otimes g(^\theta \overline{\pi })+\,^\theta \overline{s}\otimes
g(u)+(^\theta \overline{S})g\text{ ,}
\end{equation}

\noindent where 
\begin{equation}  \label{IX.2.-23}
_k\theta =\,_k\overline{\theta }_\alpha \,^\beta \cdot e_\beta \otimes
e^\alpha \text{ ,\thinspace \thinspace \thinspace \thinspace \thinspace }_k%
\overline{\theta }_\alpha \,^\beta =\overline{t}_\alpha \,^\beta -K_\alpha
\,^\beta -\overline{W}_\alpha \,^{\beta \gamma }\,_\gamma +L\cdot g_\alpha
^\beta \text{ ,}
\end{equation}
\begin{equation}  \label{IX.2.-24}
\rho _\theta =\frac 1{e^2}\cdot [g(u)](\theta )(u)\text{ , \thinspace
\thinspace \thinspace \thinspace }k=\frac 1e\cdot [g(u)](Kr)(u)\text{ ,}
\end{equation}
\begin{equation}  \label{IX.2.-25}
\begin{array}{c}
\rho _\theta =\frac 1{e^2}\cdot g_{\overline{\alpha }\overline{\beta }}\cdot
u^\beta \cdot \overline{\theta }_\gamma \,^\alpha \cdot u^{\overline{\gamma }%
}=\frac 1{e^2}\cdot \overline{\theta }_\gamma \,^\alpha \cdot u_{\overline{%
\alpha }}\cdot u^{\overline{\gamma }}= \\ 
=\frac 1{e^2}\cdot g_{\overline{i}\overline{j}}\cdot u^j\cdot \overline{%
\theta }_k\,^i\cdot u^{\overline{k}}=\frac 1{e^2}\cdot \overline{\theta }%
_k\,^i\cdot u_{\overline{i}}\cdot u^{\overline{k}}\text{ ,}
\end{array}
\end{equation}
\begin{equation}  \label{IX.2.-26}
\varepsilon _{_k\theta }=\rho _\theta +\frac 1e\cdot L\cdot k\text{ ,}
\end{equation}
\begin{equation}  \label{IX.2.-27}
^\theta \overline{\pi }=\frac 1e\cdot [g(u)](_k\theta )h^u=\,^\theta 
\overline{\pi }^\alpha \cdot e_\alpha =\,^\theta \overline{\pi }^i\cdot
\partial _i\text{ ,}
\end{equation}
\begin{equation}  \label{IX.2.-28}
^\theta \overline{\pi }^\alpha =\frac 1e\cdot \,_k\overline{\theta }_\gamma
\,^\beta \cdot u_{\overline{\beta }}\cdot h^{\overline{\gamma }\alpha }\text{
,\thinspace \thinspace \thinspace \thinspace }^\theta \overline{\pi }%
^i=\frac 1e\cdot \,_k\overline{\theta }_l\,^k\cdot u_{\overline{k}}\cdot h^{%
\overline{l}i}\text{ ,}
\end{equation}
\begin{equation}  \label{IX.2.-29}
^\theta \overline{s}=\frac 1e\cdot h^u(g)(_k\theta )(u)=\,^\theta \overline{s%
}^\alpha \cdot e_\alpha =\,^\theta \overline{s}^i\cdot \partial _i\text{ ,}
\end{equation}
\begin{equation}  \label{IX.2.-30}
^\theta \overline{s}^\alpha =\frac 1e\cdot h^{\alpha \beta }\cdot g_{%
\overline{\beta }\overline{\gamma }}\cdot \,_k\overline{\theta }_\delta
\,^\gamma \cdot u^{\overline{\delta }}\text{ ,\thinspace \thinspace
\thinspace \thinspace \thinspace }^\theta \overline{s}^i=\frac 1e\cdot
h^{ij}\cdot g_{\overline{j}\overline{k}}\cdot \,_k\overline{\theta }%
_l\,^k\cdot u^{\overline{l}}\text{ ,}
\end{equation}
\begin{equation}  \label{IX.2.-31}
^\theta \overline{S}=h^u(g)(_k\theta )h^u=\,^\theta \overline{S}\,^{\alpha
\beta }\cdot e_\alpha \otimes e_\beta =\,^\theta \overline{S}\,^{ij}\cdot
\partial _i\otimes \partial _j\text{ ,}
\end{equation}
\begin{equation}  \label{IX.2.-32}
^\theta \overline{S}\,^{\alpha \beta }=h^{\alpha \gamma }\cdot g_{\overline{%
\gamma }\overline{\delta }}\cdot \,_k\overline{\theta }_\kappa \,^\delta
\cdot h^{\overline{\kappa }\beta }\text{ ,\thinspace \thinspace \thinspace
\thinspace \thinspace \thinspace \thinspace \thinspace }^\theta \overline{S}%
\,^{ij}=h^{ik}\cdot g_{\overline{k}\overline{l}}\cdot \,_k\overline{\theta }%
_m\,^l\cdot h^{\overline{m}j}\text{ ,}
\end{equation}
\begin{equation}  \label{IX.2.-33}
\begin{array}{c}
(\theta ) \overline{g}=(\rho _\theta +\frac 1e\cdot L\cdot k)\cdot u\otimes
u-L\cdot Kr(\overline{g})+u\otimes \,^\theta \overline{\pi }+\,^\theta 
\overline{s}\otimes u+\,^\theta \overline{S}= \\ 
=\theta ^{\alpha \beta }\cdot e_\alpha \otimes e_\beta =\theta ^{ij}\cdot
\partial _i\otimes \partial _j\text{ ,}
\end{array}
\end{equation}
\begin{equation}  \label{IX.2.-34}
\theta ^{\alpha \beta }=g^{\beta \overline{\gamma }}\cdot \overline{\theta }%
_\gamma \,^\alpha =\overline{\theta }_\gamma \,^\alpha \cdot g^{\beta 
\overline{\gamma }}\text{ ,\thinspace \thinspace \thinspace \thinspace
\thinspace }\theta ^{ij}=\overline{\theta }_k\,^i\cdot g^{\overline{k}j}%
\text{ ,}
\end{equation}
\begin{equation}  \label{IX.2.-35}
\begin{array}{c}
g(\theta )=(\rho _\theta +\frac 1e\cdot L\cdot k)\cdot g(u)\otimes
g(u)-L\cdot g(Kr)+g(u)\otimes g(^\theta \overline{\pi })+ \\ 
+g(^\theta \overline{s})\otimes g(u)+g(^\theta \overline{S})g\text{ ,}
\end{array}
\end{equation}
\begin{equation}  \label{IX.2.-36}
\begin{array}{c}
g(\theta )=\theta _{\alpha \beta }\cdot e^\alpha \otimes e^\beta =\theta
_{ij}\cdot dx^i\otimes dx^j \text{ ,} \\ 
\theta _{\alpha \beta }=g_{\alpha \overline{\gamma }}\cdot \overline{\theta }%
_\beta \,^\gamma \text{ ,\thinspace \thinspace \thinspace \thinspace
\thinspace \thinspace \thinspace }\theta _{ij}=g_{i\overline{k}}\cdot 
\overline{\theta }_j\text{ }^k\text{ ,}
\end{array}
\end{equation}
\begin{equation}  \label{IX.2.-40}
\begin{array}{c}
g(Kr)=g_{\alpha \overline{\beta }}\cdot e^\alpha \otimes e^\beta =g_{i%
\overline{j}}\cdot dx^i\otimes dx^j\text{ ,\thinspace \thinspace \thinspace
\thinspace }g_{\alpha \overline{\beta }}=g_{\alpha \gamma }\cdot f^\gamma
\,_\beta =g_{\gamma \alpha }\cdot f^\gamma \,_\beta =g_{\overline{\beta }%
\alpha }\text{ ,} \\ 
g_{i\overline{j}}=g_{\overline{j}i}\text{ ,\thinspace \thinspace \thinspace
\thinspace }(Kr)\overline{g}=Kr(\overline{g})=g^{\overline{\alpha }\beta
}\cdot e_\alpha \otimes e_\beta =g^{\overline{i}j}\cdot \partial _i\otimes
\partial _j\text{ .}
\end{array}
\end{equation}

In a co-ordinate basis $\theta $, $(\theta )\overline{g}$ and $g(\theta )$
can be represented in the forms 
\begin{equation}  \label{IX.2.-46}
\overline{\theta }_i\,^j=(\rho _\theta +\frac 1e\cdot L\cdot k)\cdot
u_i\cdot u^j-L\cdot g_i^j+\,^\theta \overline{\pi }_i\cdot u^j+u_i\cdot
\,^\theta \overline{s}^j+g_{i\overline{k}}\cdot \,^\theta \overline{S}\,^{jk}%
\text{ ,}
\end{equation}
\begin{equation}  \label{IX.2.-47}
\theta ^{ij}=\overline{\theta }_k\,^i\cdot g^{\overline{k}j}=(\rho _\theta
+\frac 1e\cdot L\cdot k)\cdot u^i\cdot u^j-L\cdot g^{\overline{i}j}+u^i\cdot
\,^\theta \overline{\pi }^j+\,^\theta \overline{s}^i\cdot u^j+\,^\theta 
\overline{S}\,^{ij}\text{ ,}
\end{equation}
\begin{equation}  \label{IX.2.-48}
\theta _{ij}=g_{i\overline{k}}\cdot \overline{\theta }_j\,^k=(\rho _\theta
+\frac 1e\cdot L\cdot k)\cdot u_i\cdot u_j-L\cdot g_{i\overline{j}}+u_i\cdot
\,^\theta \overline{\pi }_j+\,^\theta \overline{s}_i\cdot u_j+\,g_{i%
\overline{k}}\cdot ^\theta \overline{S}\,^{kl}\cdot g_{\overline{l}j}\text{ ,%
}
\end{equation}

\noindent where 
\begin{equation}  \label{IX.2.-49}
u_i=g_{i\overline{j}}\cdot u^j\,\,\,\,\text{,\thinspace \thinspace
\thinspace \thinspace }^\theta \overline{\pi }_i=g_{i\overline{k}}\cdot
\,^\theta \overline{\pi }^k\text{ ,\thinspace \thinspace \thinspace
\thinspace \thinspace }^\theta \overline{s}_i=g_{i\overline{l}}\cdot
\,^\theta \overline{s}^l\text{ ,\thinspace \thinspace \thinspace }^\theta 
\overline{S}_{ij}=g_{i\overline{k}}\cdot \,^\theta \overline{S}\,^{kl}\cdot
g_{\overline{l}j}\text{ . }
\end{equation}

The symmetric energy-momentum tensor of Belinfante $_sT$ can be represented
in an analogous way by the use of the projective metrics $h^u$ and $h_u$ and
the rest mass density $\rho _T$ in the form 
\begin{equation}  \label{IX.2.-51}
_sT=(\rho _T+\frac 1e\cdot L\cdot k).u\otimes g(u)-L\cdot Kr+u\otimes g(^T%
\overline{\pi })+\,^T\overline{s}\otimes g(u)+(^T\overline{S})g\text{ ,}
\end{equation}

\noindent where 
\begin{equation}  \label{IX.2.-52}
\begin{array}{c}
_sT=\,_sT_\alpha \,^\beta \cdot e_\beta \otimes e^\alpha =\,_sT_i\,^j\cdot
\partial _j\otimes dx^i \text{ ,} \\ 
_sT=\,_{sk}T-L\cdot Kr\text{ ,\thinspace \thinspace \thinspace \thinspace
\thinspace \thinspace \thinspace }_{sk}T=\,_sT+L\cdot Kr=\mathit{T}\text{ ,}
\end{array}
\end{equation}
\begin{equation}  \label{IX.2.-53}
\varepsilon _T=\rho _T+\frac 1e\cdot L\cdot k\text{ ,}
\end{equation}
\begin{equation}  \label{IX.2.-54}
\begin{array}{c}
\rho _T=\frac 1{e^2}\cdot [g(u)](_sT)(u)=\frac 1{e^2}\cdot g_{\overline{%
\alpha }\overline{\beta }}\cdot u^\beta \cdot _sT_\gamma \,^\alpha \cdot u^{%
\overline{\gamma }}=\frac 1{e^2}\cdot _sT_\gamma \,^\alpha \cdot u_{%
\overline{\alpha }}\cdot u^{\overline{\gamma }}= \\ 
=\frac 1{e^2}\cdot g_{\overline{i}\overline{j}}\cdot u^j\cdot _sT_k\,^i\cdot
u^{\overline{k}}=\frac 1{e^2}\cdot \,_sT_k\,^i\cdot u_{\overline{i}}\cdot u^{%
\overline{k}}\text{ ,}
\end{array}
\end{equation}
\begin{equation}  \label{IX.2.-55}
^T\overline{\pi }=\frac 1e\cdot [g(u)](\mathit{T})h^u=\frac 1e\cdot
[g(u)](_{sk}T)h^u=\,^T\overline{\pi }^\alpha \cdot e_\alpha =\,^T\overline{%
\pi }^i\cdot \partial _i\text{ ,}
\end{equation}
\begin{equation}  \label{IX.2.-56}
^T\overline{\pi }^\alpha =\frac 1e\cdot \mathit{T}_\gamma \,^\beta \cdot u_{%
\overline{\beta }}\cdot h^{\overline{\gamma }\alpha }\text{ ,\thinspace
\thinspace \thinspace \thinspace }^T\overline{\pi }^i=\frac 1e\cdot \mathit{T%
}_l\,^k\cdot u_{\overline{k}}\cdot h^{\overline{l}i}\text{ ,}
\end{equation}
\begin{equation}  \label{IX.2.-57}
^T\overline{s}=\frac 1e\cdot h^u(g)(\mathit{T})(u)=\,^T\overline{s}^\alpha
\cdot e_\alpha =\,^T\overline{s}^i\cdot \partial _i\text{ ,}
\end{equation}
\begin{equation}  \label{IX.2.-58}
^T\overline{s}\,^\alpha =\frac 1e\cdot h^{\alpha \beta }\cdot g_{\overline{%
\beta }\overline{\gamma }}\cdot \mathit{T}_\delta \,^\gamma \cdot u^{%
\overline{\delta }}\text{ ,\thinspace \thinspace \thinspace \thinspace
\thinspace }^T\overline{s}^i=\frac 1e\cdot h^{ij}\cdot g_{\overline{j}%
\overline{k}}\cdot \mathit{T}_l\,^k\cdot u^{\overline{l}}\text{ ,}
\end{equation}
\begin{equation}  \label{IX.2.-59}
^T\overline{S}=h^u(g)(\mathit{T})h^u=\,^T\overline{S}\,^{\alpha \beta }\cdot
e_\alpha \otimes e_\beta =\,^T\overline{S}\,^{ij}\cdot \partial _i\otimes
\partial _j\text{ ,}
\end{equation}
\begin{equation}  \label{IX.2.-60}
^T\overline{S}\,^{\alpha \beta }=h^{\alpha \gamma }\cdot g_{\overline{\gamma 
}\overline{\delta }}\cdot \mathit{T}_\kappa \,^\delta \cdot h^{\overline{%
\kappa }\beta }\text{ ,\thinspace \thinspace \thinspace \thinspace
\thinspace \thinspace \thinspace \thinspace }^T\overline{S}%
\,^{ij}=h^{ik}\cdot g_{\overline{k}\overline{l}}\cdot \mathit{T}_m\,^l\cdot
h^{\overline{m}j}\text{ ,}
\end{equation}
\begin{equation}  \label{IX.2.-61}
\begin{array}{c}
(_sT) \overline{g}=(\rho _T+\frac 1e\cdot L\cdot k)\cdot u\otimes u-L\cdot
Kr(\overline{g})+u\otimes \,^T\overline{\pi }+\,^T\overline{s}\otimes u+\,^T%
\overline{S}= \\ 
=\,_sT^{\alpha \beta }\cdot e_\alpha \otimes e_\beta =\,_sT^{ij}\cdot
\partial _i\otimes \partial _j\text{ ,}
\end{array}
\end{equation}
\begin{equation}  \label{IX.2.-62}
_sT^{\alpha \beta }=g^{\beta \overline{\gamma }}\cdot \,_sT_\gamma \,^\alpha
=\,_sT_\gamma \,^\alpha \cdot g^{\beta \overline{\gamma }}\text{ ,\thinspace
\thinspace \thinspace \thinspace \thinspace }_sT^{ij}=\,_sT_k\,^i\cdot g^{%
\overline{k}j}\text{ ,}
\end{equation}
\begin{equation}  \label{IX.2.-63}
\begin{array}{c}
g(_sT)=(\rho _T+\frac 1e\cdot L\cdot k)\cdot g(u)\otimes g(u)-L\cdot
g(Kr)+g(u)\otimes g(^T \overline{\pi })+ \\ 
+g(^T\overline{s})\otimes g(u)+g(^T\overline{S})g\text{ ,}
\end{array}
\end{equation}
\begin{equation}  \label{IX.2.-64}
\begin{array}{c}
g(_sT)=\,_sT_{\alpha \beta }\cdot e^\alpha \otimes e^\beta =\,_sT_{ij}\cdot
dx^i\otimes dx^j \text{ ,} \\ 
_sT_{\alpha \beta }=g_{\alpha \overline{\gamma }}\cdot \,_sT_\beta \,^\gamma 
\text{ ,\thinspace \thinspace \thinspace \thinspace \thinspace \thinspace
\thinspace }_sT_{ij}=g_{i\overline{k}}\cdot \,_sT_j\text{ }^k\text{ .}
\end{array}
\end{equation}

In a co-ordinate basis $_sT$, $(_sT)\overline{g}$ and $g(_sT)$ can be
represented in the forms 
\begin{equation}  \label{IX.2.-68}
_sT_i\,^j=(\rho _T+\frac 1e\cdot L\cdot k)\cdot u_i\cdot u^j-L\cdot
g_i^j+\,^T\overline{\pi }_i\cdot u^j+u_i\cdot ^T\overline{s}^j+g_{i\overline{%
k}}\cdot \,^T\overline{S}\,^{jk}\text{ ,}
\end{equation}
\begin{equation}  \label{IX.2.-69}
_sT^{ij}=\,_sT_k\,^i\cdot g^{\overline{k}j}=(\rho _T+\frac 1e\cdot L\cdot
k)\cdot u^i\cdot u^j-L\cdot g^{\overline{i}j}+u^i\cdot \,^T\overline{\pi }%
^j+\,^T\overline{s}^i\cdot u^j+\,^T\overline{S}\,^{ij}\text{ ,}
\end{equation}
\begin{equation}  \label{IX.2.-70}
_sT_{ij}=g_{i\overline{k}}\cdot \,_sT_j\,^k=(\rho _T+\frac 1e\cdot L\cdot
k)\cdot u_i\cdot u_j-L\cdot g_{i\overline{j}}+u_i\cdot \,^T\overline{\pi }%
_j+\,^T\overline{s}_i\cdot u_j+\,g_{i\overline{k}}\cdot \,^T\overline{S}%
\,^{kl}\cdot g_{\overline{l}j}\text{ ,}
\end{equation}

\noindent where 
\[
^T\overline{\pi }_i=g_{i\overline{k}}\cdot \,^T\overline{\pi }^k\text{
,\thinspace \thinspace \thinspace \thinspace \thinspace }^T\overline{s}%
_i=g_{i\overline{l}}\cdot \,^T\overline{s}^l\text{ ,\thinspace \thinspace
\thinspace }^T\overline{S}_{ij}=g_{i\overline{k}}\cdot \,^T\overline{S}%
\,^{kl}\cdot g_{\overline{l}j}\text{ . } 
\]

The variational energy-momentum tensor of Euler-Lagrange $Q$ can be
represented in the standard manner by the use of the projective metrics $h^u$%
, $h_u$ and the rest mass density $\rho _Q$ in the form

\begin{equation}  \label{IX.2.-71}
Q=-\rho _Q\cdot u\otimes g(u)-u\otimes g(^Q\pi )-\,^Qs\otimes g(u)-(^QS)g%
\text{ ,}
\end{equation}

\noindent where 
\begin{equation}  \label{IX.2.-72}
\rho _Q=-\,\,\frac 1{e^2}\cdot [g(u)](Q)(u)\text{ , }
\end{equation}
\begin{equation}  \label{IX.2.-72a}
\begin{array}{c}
\rho _Q=-\,\,\frac 1{e^2}\cdot g_{\overline{\alpha }\overline{\beta }}\cdot
u^\beta \cdot \overline{Q}_\gamma \,^\alpha \cdot u^{\overline{\gamma }%
}=-\,\frac 1{e^2}\cdot \overline{Q}_\gamma \,^\alpha \cdot u_{\overline{%
\alpha }}\cdot u^{\overline{\gamma }}= \\ 
=-\,\frac 1{e^2}\cdot g_{\overline{i}\overline{j}}\cdot u^j\cdot \overline{Q}%
_k\,^i\cdot u^{\overline{k}}=-\,\frac 1{e^2}\cdot \overline{Q}_k\,^i\cdot u_{%
\overline{i}}\cdot u^{\overline{k}}\text{ ,}
\end{array}
\end{equation}
\begin{equation}  \label{IX.2.-73}
^Q\pi =-\frac 1e\cdot [g(u)](Q)h^u=\,^Q\pi ^\alpha \cdot e_\alpha =\,^Q\pi
^i\cdot \partial _i\text{ ,}
\end{equation}
\begin{equation}  \label{IX.2.-74}
^Q\pi ^\alpha =-\frac 1e\cdot \overline{Q}_\gamma \,^\beta \cdot u_{%
\overline{\beta }}\cdot h^{\overline{\gamma }\alpha }\text{ ,\thinspace
\thinspace \thinspace \thinspace }^Q\pi ^i=-\frac 1e\cdot \overline{Q}%
_l\,^k\cdot u_{\overline{k}}\cdot h^{\overline{l}i}\text{ ,}
\end{equation}
\begin{equation}  \label{IX.2.-75}
^Qs=-\frac 1e\cdot h^u(g)(Q)(u)=\,^Qs^\alpha \cdot e_\alpha =\,^Qs^i\cdot
\partial _i\text{ ,}
\end{equation}
\begin{equation}  \label{IX.2.-76}
^Qs^\alpha =-\frac 1e\cdot h^{\alpha \beta }\cdot g_{\overline{\beta }%
\overline{\gamma }}\cdot \overline{Q}_\delta \,^\gamma \cdot u^{\overline{%
\delta }}\text{ ,\thinspace \thinspace \thinspace \thinspace \thinspace }%
^Qs^i=-\frac 1e\cdot h^{ij}\cdot g_{\overline{j}\overline{k}}\cdot \overline{%
Q}_l\,^k\cdot u^{\overline{l}}\text{ ,}
\end{equation}
\begin{equation}  \label{IX.2.-77}
^QS=-h^u(g)(Q)h^u=\,^QS^{\alpha \beta }\cdot e_\alpha \otimes e_\beta
=\,^QS^{ij}\cdot \partial _i\otimes \partial _j\text{ ,}
\end{equation}
\begin{equation}  \label{IX.2.-78}
^QS^{\alpha \beta }=-h^{\alpha \gamma }\cdot g_{\overline{\gamma }\overline{%
\delta }}\cdot \overline{Q}_\kappa \,^\delta \cdot h^{\overline{\kappa }%
\beta }\text{ ,\thinspace \thinspace \thinspace \thinspace \thinspace
\thinspace \thinspace \thinspace }^QS^{ij}=-h^{ik}\cdot g_{\overline{k}%
\overline{l}}\cdot \overline{Q}_m\,^l\cdot h^{\overline{m}j}\text{ ,}
\end{equation}
\begin{equation}  \label{IX.2.-82}
\begin{array}{c}
(Q) \overline{g}=-\,\rho _Q\cdot u\otimes u-u\otimes \,^Q\pi -\,^Qs\otimes
u-\,^QS= \\ 
=Q^{\alpha \beta }\cdot e_\alpha \otimes e_\beta =Q^{ij}\cdot \partial
_i\otimes \partial _j\text{ ,}
\end{array}
\end{equation}
\[
Q^{\alpha \beta }=g^{\beta \overline{\gamma }}\cdot \overline{Q}_\gamma
\,^\alpha =\overline{Q}_\gamma \,^\alpha \cdot g^{\beta \overline{\gamma }}%
\text{ ,\thinspace \thinspace \thinspace \thinspace \thinspace }Q^{ij}=%
\overline{Q}_k\,^i\cdot g^{\overline{k}j}\text{ ,} 
\]
\begin{equation}  \label{IX.2.-79}
g(Q)=-\rho _Q\cdot g(u)\otimes g(u)-g(u)\otimes g(^Q\pi )-g(^Qs)\otimes
g(u)-g(^QS)g\text{ ,}
\end{equation}
\begin{equation}  \label{IX.2.-81}
g(Q)=Q_{\alpha \beta }\cdot e^\alpha \otimes e^\beta =Q_{ij}\cdot
dx^i\otimes dx^j\text{ ,\thinspace \thinspace \thinspace }Q_{\alpha \beta
}=g_{\alpha \overline{\gamma }}\cdot \overline{Q}_\beta \,^\gamma \text{
,\thinspace \thinspace \thinspace \thinspace \thinspace \thinspace
\thinspace }Q_{ij}=g_{i\overline{k}}\cdot \overline{Q}_j\text{ }^k\text{ .}
\end{equation}

In a co-ordinate basis $Q$, $(Q)\overline{g}$ and $g(Q)$ can be represented
in the forms 
\begin{equation}  \label{IX.2.-86}
\overline{Q}_i\,^j=-\rho _Q\cdot u_i\cdot u^j-\,^Q\pi _i\cdot u^j-u_i\cdot
\,^Qs^j-g_{i\overline{k}}\cdot \,^QS^{jk}\text{ ,}
\end{equation}

\begin{equation}  \label{IX.2.-87}
Q^{ij}=\overline{Q}_k\,^i\cdot g^{\overline{k}j}=-\rho _Q\cdot u^i\cdot
u^j-u^i\cdot \,^Q\pi ^j-\,^Qs^i\cdot u^j-\,^QS^{ij}\text{ ,}
\end{equation}

\begin{equation}  \label{IX.2.-88}
Q_{ij}=g_{i\overline{k}}\cdot \overline{Q}_j\,^k=-\rho _Q\cdot u_i\cdot
u_j-u_i\cdot \,^Q\pi _j-\,^Qs_i\cdot u_j-\,g_{i\overline{k}}\cdot
\,^QS^{kl}\cdot g_{\overline{l}j}\text{ ,}
\end{equation}

\noindent where 
\[
\text{\thinspace }^Q\pi _i=g_{i\overline{k}}\cdot \,^Q\pi ^k\text{
,\thinspace \thinspace \thinspace \thinspace \thinspace }^Qs_i=g_{i\overline{%
l}}\cdot \,^Qs^l\text{ ,\thinspace \thinspace \thinspace }^QS_{ij}=g_{i%
\overline{k}}\cdot \,^QS^{kl}\cdot g_{\overline{l}j}\text{ . } 
\]

The introduced abbreviations for the different projections of the
energy-momentum tensors have their analogous forms in $V_3$- and $V_4$%
-spaces, where their physical interpretations have been proposed \cite
{Landau-1}, \cite{Schmutzer-1} (S.383-385). The stress tensor in $V_3$%
-spaces has been generalized to the energy-momentum tensor $_sT$ in $V_4$%
-spaces. The viscosity stress tensor $_{sk}T$ appears as the tensor $\mathit{%
T}$ in the structure of the symmetric energy-momentum tensor of Belinfante $%
_sT$.

On the analogy of the physical interpretation of the different projections,
the following definitions can be proposed for the quantities in the
representations of the different energy-momentum tensors:

\textit{A. Generalized canonical energy-momentum tensor of the type 1}.
........ $\theta $

(a) Generalized viscous energy-momentum tensor of the type 1. ............ $%
_k\theta $

(b) Rest mass density of the generalized canonical

energy-momentum tensor $\theta $.
....................................................................$\rho
_\theta $

(c) Conductive momentum density of the generalized canonical

energy-momentum tensor $\theta $.
...................................................................$^\theta
\pi $

(d) Conductive energy flux density of the generalized canonical

energy-momentum tensor $\theta $.
..................................................................$e\cdot
\,^\theta s$

(e) Stress tensor of the generalized canonical

energy-momentum tensor $\theta $.
....................................................................$^\theta
S$

\textit{B. Symmetric energy-momentum tensor of Belinfante of the type 1.}
..... $_sT$

(a) Symmetric viscous energy-momentum tensor of the type 1. ...............$%
...\mathit{T}$

(b) Rest mass density of the symmetric energy-momentum

tensor of Belinfante $_sT$.
...........................................................................$%
\rho _T$

(c) Conductive momentum density of the symmetric

energy-momentum tensor of Belinfante $_sT$.
..............................................$^T\pi $

(d) Conductive energy flux density of the symmetric

energy-momentum tensor of Belinfante $_sT$.
............................................$e\cdot \,^Ts$

(e) Stress tensor of the symmetric energy-momentum

tensor of Belinfante $_sT$.
..........................................................................$%
^TS$

\textit{C. Variational (active) energy-momentum tensor of Euler-Lagrange.}
.. $Q $

(a) Rest mass density of the variational energy-momentum

tensor of Euler-Lagrange $Q$%
..................................................................... $\rho
_Q$

(b) Conductive momentum density of the variational

energy-momentum tensor of Euler-Lagrange $Q$%
.........................................$^Q\pi $

(c) Conductive energy flux density of the variational

energy-momentum tensor of Euler-Lagrange $Q$%
....................................... $e\cdot \,^Qs$

(d) Stress tensor of the variational energy-momentum

tensor of Euler-Lagrange $Q$.
....................................................................$^QS$

The projections of the energy-momentum tensors have properties which are due
to their construction, the orthogonality of the projective metrics $h_u$ and 
$h^u$ correspondingly to the vector fields $u$ and $g(u)$ [$h_u(u)=0$%
,\thinspace \thinspace \thinspace \thinspace \thinspace \thinspace $%
h^u[g(u)]=0$] 
\begin{equation}  \label{IX.2.-89}
g(u,^\theta \overline{\pi })=g(^\theta \overline{\pi },u)=0\text{
,\thinspace \thinspace \thinspace \thinspace }g(u,\,^T\overline{\pi })=0%
\text{ ,\thinspace \thinspace \thinspace \thinspace }g(u,\,^Q\pi )=0\text{ ,}
\end{equation}
\begin{equation}  \label{IX.2.-90}
g(u,^\theta \overline{s})=g(^\theta \overline{s},u)=0\text{ ,\thinspace
\thinspace \thinspace \thinspace }g(u,^T\overline{s})=0\text{ ,\thinspace
\thinspace \thinspace \thinspace }g(u,^Qs)=0\text{ ,}
\end{equation}
\begin{equation}  \label{IX.2.-91}
\begin{array}{c}
g(u)(^\theta \overline{S})=0\text{, }(^\theta \overline{S})g(u)=0\text{
,\thinspace \thinspace \thinspace \thinspace }g(u)(^T\overline{S})=0\text{
,\thinspace \thinspace \thinspace \thinspace \thinspace }(^T\overline{S}%
)g(u)=0\text{ ,\thinspace \thinspace } \\ 
\text{\thinspace \thinspace }g(u)(^QS)=0\text{ , \thinspace \thinspace
\thinspace \thinspace \thinspace \thinspace }(^QS)g(u)=0\text{ .\thinspace
\thinspace }
\end{array}
\end{equation}

From the properties of the different projections it follows that the
conductive momentum density $\pi $ (or $\overline{\pi }$) is a contravariant
vector field orthogonal to the vector field $u$. The conductive energy flux
density $e\cdot s$ (or $e\cdot \overline{s}$) is also a contravariant vector
field orthogonal to $u$. The stress tensor $S$ (or $\overline{S}$) is
orthogonal to $u$ independently of the side of the projection by means of
the vector field $u$.

The second covariant Noether identity $\theta -\,_sT\equiv Q$ can be written
by the use of the projections of the energy-momentum tensors in the form 
\begin{equation}  \label{IX.2.-92}
\begin{array}{c}
(\rho _\theta -\rho _T+\rho _Q)\cdot u\otimes g(u)+u\otimes g(^\theta 
\overline{\pi }-\,^T\overline{\pi }+\,^Q\pi )+ \\ 
+\,(^\theta \overline{s}-\,^T\overline{s}+\,^Qs)\otimes g(u)+(^\theta 
\overline{S}-\,^T\overline{S}+\,^QS)g\equiv 0\text{ .}
\end{array}
\end{equation}

After contraction of the last expression consistently with $u$ and $%
\overline{g}$ and taking into account the properties (\ref{IX.2.-89}) $\div $
(\ref{IX.2.-91}) the second covariant Noether identity disintegrates in
identities for the different projections of the energy-momentum tensors 
\begin{equation}  \label{IX.2.-94}
\rho _\theta \equiv \rho _T-\rho _Q\text{ ,\thinspace \thinspace \thinspace
\thinspace }^\theta \overline{\pi }\equiv \,^T\overline{\pi }-\,^Q\pi \text{
,\thinspace \thinspace \thinspace \thinspace }^\theta \overline{s}\equiv \,^T%
\overline{s}-\,^Qs\text{ , }^\theta \overline{S}\equiv \,^T\overline{S}-\,^QS%
\text{ .}
\end{equation}

If the covariant Euler-Lagrange equations of the type $\delta _vL/\delta
V^A\,_B=0$ are fulfilled for the non-metric tensor fields of a Lagrangian
system and $_gQ=0$, then the variational energy-momentum of Euler-Lagrange $%
Q=\,_vQ+\,_gQ$ is equal to zero. This fact leads to vanishing the invariant
projections of $Q$ ($\rho _Q=0$, $^Q\pi =0$, $^Qs=0$, $^QS=0$). The equality
which follows between $\theta $ and $_sT$ has as corollaries the identities 
\begin{equation}  \label{IX.2.-95}
\rho _\theta \equiv \rho _T\text{ ,\thinspace \thinspace \thinspace
\thinspace \thinspace \thinspace \thinspace \thinspace \thinspace \thinspace
\thinspace \thinspace \thinspace \thinspace \thinspace \thinspace }^\theta 
\overline{\pi }\equiv \,^T\overline{\pi }\text{ ,\thinspace \thinspace
\thinspace \thinspace \thinspace \thinspace \thinspace \thinspace \thinspace
\thinspace \thinspace \thinspace \thinspace \thinspace \thinspace \thinspace 
}^\theta \overline{s}\equiv \,^T\overline{s}\text{ ,\thinspace \thinspace
\thinspace \thinspace \thinspace \thinspace \thinspace \thinspace \thinspace
\thinspace \thinspace \thinspace \thinspace \thinspace \thinspace \thinspace
\thinspace \thinspace \thinspace \thinspace }^\theta \overline{S}\equiv \,^T%
\overline{S}\text{ .\thinspace }
\end{equation}

From the first identity ($\rho _\theta \equiv \rho _T$) and the identity (%
\ref{IX.2.-94}) for $\rho $ , it follows that the covariant Euler-Lagrange
equations of the type $\delta _vL/\delta V^A\,_B=0$ for non-metric fields $V$
and $_gQ=0$ appear as sufficient conditions for the unique determination of
the notion of rest mass density $\rho $ for a given Lagrangian system.

\begin{proposition}
The necessary and sufficient condition for the equality 
\begin{equation}
\rho _\theta =\rho _T  \label{IX.2.-96}
\end{equation}
\end{proposition}

\noindent \textit{is the condition} $\rho _Q=0$.

Proof. It follows immediately from the first identity in (\ref{IX.2.-94}).

The condition $\rho _Q\neq 0$ leads to the violation of the unique
determination of the notion of rest mass density and to the appearance of
three different notions of rest mass density corresponding to the three
different energy-momentum tensors for a Lagrangian system. Therefore, the
violation of the covariant Euler-Lagrange equations $\delta _vL/\delta
V^A\,_B=0$ for the non-metric tensor fields or the existence of metric
tensor fields in a Lagrangian density with $_gQ\neq 0$ induce a new rest
mass density (a new rest mass respectively) for which the identity (\ref
{IX.2.-94}) is fulfilled.

The identity $\rho _\theta \equiv \rho _T-\rho _Q$ can be related to the
physical hypotheses about the inertial, passive and active gravitational
rest mass densities in models for describing the gravitational interaction.
To every energy-momentum tensor a non-null rest mass density corresponds.
The existence of the variational energy-momentum tensor of Euler-Lagrange is
connected with the existence of the gravitational interaction in a
Lagrangian system in Einstein's theory of gravitation \cite{Manoff-00} and
therefore, with the existence of a non-null active gravitational rest mass
density. When a Lagrangian system does not interact gravitationaly, the
active gravitational rest mass density is equal to zero and the principle of
equivalence between the inertial and the passive rest mass density is
fulfilled \cite{Manoff-1} - \cite{Manoff-4}.

From the second covariant Noether identity of the type 1. by means of the
relations

\[
\overline{G}=g(G)\overline{g}\text{ ,\thinspace \thinspace \thinspace
\thinspace \thinspace \thinspace \thinspace \thinspace \thinspace \thinspace
\thinspace \thinspace \thinspace \thinspace \thinspace \thinspace \thinspace
\thinspace \thinspace \thinspace \thinspace \thinspace }G=\overline{g}(%
\overline{G})g\text{ ,} 
\]
one can find the second covariant Noether identity of the type 2. for the
energy-momentum tensors of the type 2. in the form 
\begin{equation}  \label{IX.2.-98}
\overline{\theta }-\,_s\overline{T}\equiv \overline{Q}\text{ ,}
\end{equation}

\noindent where 
\begin{equation}  \label{IX.2.-99}
\overline{\theta }=g(\theta )\overline{g}\text{ ,\thinspace \thinspace
\thinspace \thinspace \thinspace \thinspace \thinspace \thinspace \thinspace
\thinspace \thinspace \thinspace \thinspace }_s\overline{T}=g(_sT)\overline{g%
}\text{ ,\thinspace \thinspace \thinspace \thinspace \thinspace \thinspace
\thinspace \thinspace \thinspace \thinspace \thinspace \thinspace \thinspace
\thinspace \thinspace \thinspace \thinspace }\overline{Q}=g(Q)\overline{g}%
\text{ .}
\end{equation}

The invariant representation of the energy-momentum tensors by means of the
projective metrics $h^u$, $h_u$ and the rest mass density allows a
comparison of these tensors with the well known energy-momentum tensors from
the continuum media mechanics (for instance, with the energy-momentum tensor
of an ideal liquid in $V_4$-spaces: 
\begin{equation}  \label{IX.2.-121}
_sT_i\,^j=(\rho +\frac 1e\cdot p)\cdot u_i\cdot u^j-p\cdot g_i^j\text{
,\thinspace \thinspace \thinspace \thinspace \thinspace \thinspace
\thinspace \thinspace \thinspace \thinspace \thinspace \thinspace \thinspace 
}e=\text{const. }\neq 0\text{, \thinspace \thinspace \thinspace }k=1\text{ ).%
}
\end{equation}

It follows from the comparison that the Lagrangian invariant $L$ can be
interpreted as the pressure $p=L$ characterizing the Lagrangian system. This
possibility for an other physical interpretation than the usual one (in the
mechanics $L$ is interpreted as the difference between the kinetic and the
potential energy) allows a description of Lagrangian systems on the basis of
phenomenological investigations determining the dependence of the pressure
on other dynamical characteristics of the system. If these relations are
given, then by the use of the method of Lagrangians with covariant
derivatives (MLCD) the corresponding covariant Euler-Lagrange equations can
be found as well as the energy-momentum tensors.

The use of a contravariant non-isotropic (non-null) vector field $u$ and its
corresponding projective metrics $h^u$\thinspace and $h_u$ is analogous to
the application of a non-isotropic (time-like) vector field in the s. c. 
\textit{monad formalism} [$(3+1)\,$-\textit{decomposition}] in $V_4$-spaces
for description of dynamical systems in Einstein's theory of gravitation
(ETG) \cite{Eckart} - \cite{Vladimirov}, \cite{Schmutzer-1}, \cite
{Schmutzer-2}.

The contravariant time-like vector field has been interpreted as a
tangential vector field at the world line of an observer determining the
frame of reference (the reference frame) in the space-time. By the use of
this reference frame a given physical system is observed and described. The
characteristics of the vector field determine the properties of the
reference frame. Moreover, the vector field is assumed to be an absolute
element in the scheme for describing the physical processes, i. e. the
vector field is not an element of the model of the physical system. It is
introduced as \textit{a priory} given vector field which does not depend on
the Lagrangian system. In fact, the physical interpretation of the
contravariant non-isotropic vector field $u$ can be related to two different
approaches analogous to the method of Lagrange and the method of Euler in
describing the motion of liquids in the hydrodynamics \cite{Pavlenko}.

In the method of Lagrange, the object with the considered motion appears as
a point (particle) of the liquid. The motion of this point is given by means
of equations for the vector field $u$ interpreted as the velocity of the
particle. The solutions of these equations give the trajectories of the
points in the liquid as basic characteristics of the physical system. In
this case, the vector field $u$ appears as an element of the model of the
system. It is connected with the motions of the system's elements.
Therefore, a Lagrangian system (and respectively its Lagrangian density)
could contain as an internal characteristic a contravariant vector field $u$
obeying equations of the type of the Euler-Lagrange equations and describing
the evolution of the system.

In the method of Euler, the object with the considered motion appears as the
model of the continuum media. Instead of the investigation of the motion of
every (fixed by its velocity and position) point (particle), the kinematic
characteristics in every immovable point in the space are considered as well
as the change of these characteristics after moving on to an other space
point. The motion is assumed to be described if the vector field $u$ is
considered as a given (or known) velocity vector field.

The $(n-1)+1$ decomposition (monad formalism) can be related to the method
of Euler or to the method of Lagrange:

(a) \textit{Method of Euler}. The vector field $u$ is interpreted as the
velocity vector field of an observer who describes a physical system with
respect to his vector field (his velocity). This physical system is
characterized by means of a Lagrangian system (Lagrangian density).

The motion of the observer (his velocity vector field) is given
independently of the motion of the considered Lagrangian system.

(b) \textit{Method of Lagrange}. The vector field $u$ is interpreted as the
velocity vector field of a continuum media with a co-moving with it
observer. The last assumption means that the velocity vector field of the
observer is identical with the velocity vector field of the media where he
is situated.

The motion of the observer (his velocity vector field) is determined by the
characteristics of the (Lagrangian) system. Its velocity vector field is, on
the other side, determined by the dynamical characteristics of the system by
means of equations of the type of the Euler-Lagrange equations.

\section{Energy-momentum tensors and the momentum density}

In the first section of this paper the notion of momentum density of an
energy-momentum tensor $G$ has been introduced as 
\[
p_G=\rho _G\cdot u+\,^G\pi \text{ ,} 
\]

\noindent where $\rho _G.u$ is the \textit{convective momentum density} of
the energy-momentum tensor $G$; $^G\pi $ is the \textit{conductive momentum
density} of the energy-momentum tensor $G$. The tensor field $G$ can be
represented by means of the projective metrics $h^u$ and $h_u$ in the form 
\[
G=u\otimes g(p_G)+\,^Gs\otimes g(u)+(^GS)g\text{ .} 
\]

At the same time the relations are fulfilled 
\begin{equation}  \label{IX.3.-2}
(G)\overline{g}=u\otimes p_G+\,^Gs\otimes u+\,^GS=G^{\alpha \beta }\cdot
e_\alpha \otimes e_\beta =G^{ij}\cdot \partial _i\otimes \partial _j\text{ ,}
\end{equation}

\begin{equation}  \label{IX.3.-3}
g(G)=g(u)\otimes g(p_G)+g(^Gs)\otimes g(u)+g(^GS)g\text{ ,}
\end{equation}

\begin{equation}  \label{IX.3.-4}
p_G=\frac 1e\cdot [g(u)](G)(\overline{g})\text{ .}
\end{equation}

The Kronecker tensor can be represented in the form 
\begin{equation}  \label{IX.3.-5}
Kr=u\otimes g(p_{Kr})+\,^{Kr}s\otimes g(u)+(^{Kr}S)g\text{ ,}
\end{equation}

\noindent where 
\begin{equation}  \label{IX.3.-6}
p_{Kr}=\frac 1e\cdot k\cdot u+\,^{Kr}\pi ,\,\,\,\,\,\,L\cdot Kr=u\otimes
g(L\cdot p_{Kr})+L\cdot \,^{Kr}s\otimes g(u)+(L\cdot \,^{Kr}S)g\text{ .}
\end{equation}

The generalized canonical energy-momentum tensor $\theta $ will have then
the form 
\begin{equation}  \label{IX.3.-7}
\theta =u\otimes g(p_\theta )+\,^\theta s\otimes g(u)+(^\theta S)g\text{ ,}
\end{equation}

\noindent where 
\begin{equation}  \label{IX.3.-8}
p_\theta =\rho _\theta \cdot u+\,^\theta \pi =\rho _\theta \cdot u+\,^\theta 
\overline{\pi }-L\cdot \,^{Kr}\pi \text{ .}
\end{equation}

In a co-ordinate basis $\theta $, $(\theta )\overline{g}$ and $g(\theta )$
can be written in the forms 
\begin{equation}  \label{IX.3.-12}
\overline{\theta }_i\,^j=p_{\theta i}\cdot u^j+u_i\cdot \,^\theta s^j+g_{i%
\overline{k}}\cdot \,^\theta S^{jk}\text{ ,}
\end{equation}

\[
p_\theta =p_\theta ^i\cdot \partial _i\text{ ,\thinspace \thinspace
\thinspace \thinspace }p_{\theta j}=g_{j\overline{k}}\cdot p_\theta ^k\text{
,} 
\]

\begin{equation}  \label{IX.3.-13}
\theta ^{ij}=\overline{\theta }_k\,^i\cdot g^{\overline{k}j}=u^i\cdot
p_\theta ^j+\,^\theta s^i\cdot u^j+\,^\theta S^{ij}\text{ ,}
\end{equation}

\begin{equation}  \label{IX.3.-14}
\theta _{ij}=g_{i\overline{k}}\cdot \overline{\theta }_j\,^k=u_i\cdot
p_{\theta j}+\,^\theta s_i\cdot u_j+\,^\theta S_{ij}\text{ .}
\end{equation}

The symmetric energy-momentum tensor of Belinfante $_sT$ can be represented
by means of the projective metrics $h^u$, $h_u$ and the momentum density $%
p_T $ in the form 
\begin{equation}  \label{IX.3.-15}
\begin{array}{c}
_sT=u\otimes g(p_T)+\,^Ts\otimes g(u)+(^TS)g= \\ 
=\,_sT_\alpha \,^\beta \cdot e_\beta \otimes e^\alpha =\,_sT_i\,^j\cdot
\partial _j\otimes dx^i\text{ ,}
\end{array}
\end{equation}

\noindent where 
\begin{equation}  \label{IX.3.-16}
p_T=\rho _T\cdot u+\,^T\pi =\rho _T\cdot u+\,^T\overline{\pi }-L\cdot \text{%
\thinspace }^{Kr}\pi \text{ ,}
\end{equation}

\begin{equation}  \label{IX.3.-17}
(_sT)\overline{g}=u\otimes p_T+\,^Ts\otimes u+\,^TS=\,_sT^{\alpha \beta
}\cdot e_\alpha \otimes e_\beta =\,_sT^{ij}\cdot \partial _i\otimes \partial
_j\text{ ,}
\end{equation}

\begin{equation}  \label{IX.3.-18}
\begin{array}{c}
g(_sT)=g(u)\otimes g(p_T)+g(^Ts)\otimes g(u)+g(^TS)g= \\ 
=\,_sT_{\alpha \beta }\cdot e^\alpha \otimes e^\beta =\,_sT_{ij}\cdot
dx^i\otimes dx^j\text{ ,}
\end{array}
\end{equation}

\begin{equation}  \label{IX.3.-19}
g(u,p_T)=\rho _T\cdot e\text{ ,\thinspace \thinspace \thinspace \thinspace
\thinspace \thinspace \thinspace \thinspace \thinspace \thinspace \thinspace 
}g(u,\,^T\pi )=0\text{ ,\thinspace \thinspace \thinspace \thinspace
\thinspace \thinspace \thinspace \thinspace \thinspace \thinspace \thinspace 
}h_u(p_T)=h_u(^T\pi )\text{ , }
\end{equation}

\begin{equation}  \label{IX.3.-21}
h^u(h_u)=h^u(g)=\overline{g}(h_u)\text{ ,\thinspace \thinspace \thinspace
\thinspace \thinspace }h_u(^G\pi )=(^G\pi )h_u=g(^G\pi )=(^G\pi )g\text{ ,}
\end{equation}

\begin{equation}  \label{IX.3.-22}
h_u(p_G)=g(^G\pi )\text{ ,\thinspace \thinspace \thinspace \thinspace
\thinspace }^G\pi =\overline{g}[h_u(p_G)]\text{ ,}
\end{equation}

\[
g(^T\pi )=h_u(p_T)\text{ ,\thinspace \thinspace \thinspace \thinspace }^T\pi
=\overline{g}[h_u(p_T)]\text{ .} 
\]

In a co-ordinate basis $_sT$, $(_sT)\overline{g}$ and $g(_sT)$ can be
written in the forms 
\begin{equation}  \label{IX.3.-27}
_sT_i\,^j=p_{Ti}\cdot u^j+u_i\cdot \,^Ts^j+g_{i\overline{k}}\cdot \,^TS^{jk}%
\text{ ,}
\end{equation}

\[
p_T=p_T^i\cdot \partial _i\text{ ,\thinspace \thinspace \thinspace
\thinspace }p_{Tj}=g_{j\overline{k}}\cdot p_T^k\text{ ,} 
\]

\begin{equation}  \label{IX.3.-28}
_sT^{ij}=\,_sT_k\,^i\cdot g^{\overline{k}j}=u^i\cdot p_T^j+\,^Ts^i\cdot
u^j+\,^TS^{ij}\text{ ,}
\end{equation}

\begin{equation}  \label{IX.3.-29}
_sT_{ij}=g_{i\overline{k}}\cdot \,_sT_j\,^k=u_i\cdot p_{Tj}+\,^Ts_i\cdot
u_j+\,^TS_{ij}\text{ .}
\end{equation}

The variational energy-momentum tensor of Euler-Lagrange $Q$ can be
represented by means of the projective metrics $h^u$, $h_u$ and the momentum
density $p_Q$ in the forms 
\begin{equation}  \label{IX.3.-30}
\begin{array}{c}
Q=-\,\,\,u\otimes g(p_Q)-\,^Qs\otimes g(u)-(^QS)g= \\ 
=\,\overline{Q}_\alpha \,^\beta \cdot e_\beta \otimes e^\alpha =\,\overline{Q%
}_i\,^j\cdot \partial _j\otimes dx^i\text{ ,}
\end{array}
\end{equation}

\begin{equation}  \label{IX.3.-31}
(Q)\overline{g}=-\,\,\,u\otimes p_Q-\,^Qs\otimes u-\,^QS=Q^{\alpha \beta
}\cdot e_\alpha \otimes e_\beta =\,Q^{ij}\cdot \partial _i\otimes \partial _j%
\text{ ,}
\end{equation}

\begin{equation}  \label{IX.3.-32}
\begin{array}{c}
g(Q)=-\,\,\,g(u)\otimes g(p_Q)-g(^Qs)\otimes g(u)-g(^QS)g= \\ 
=\,Q_{\alpha \beta }\cdot e^\alpha \otimes e^\beta =Q_{ij}\cdot dx^i\otimes
dx^j\text{ ,}
\end{array}
\end{equation}

\noindent where 
\begin{equation}  \label{IX.3.-33}
\begin{array}{c}
h_u(^Q\pi )=g(^Q\pi ) \text{ ,\thinspace \thinspace \thinspace \thinspace }%
g(u,^Q\pi )=0\text{ ,\thinspace \thinspace \thinspace \thinspace \thinspace }%
h_u(p_Q)=g(^Q\pi )\text{ ,} \\ 
p_Q=\rho _Q\cdot u+\,^Q\pi \text{ ,\thinspace \thinspace \thinspace
\thinspace \thinspace }^Q\pi =\overline{g}[h_u(p_Q)]\text{ , \thinspace
\thinspace \thinspace \thinspace }h_u(^Qs)=g(^Qs)\text{ ,} \\ 
h_u(h^u)=h^u(h_u)=h_u(\overline{g})=g(h^u)\text{ .}
\end{array}
\end{equation}

In a co-ordinate basis $Q$, $(Q)\overline{g}$ and $g(Q)$ will have the forms 
\begin{equation}  \label{IX.3.-37}
\overline{Q}_i\,^j=-\,\,\,\,p_{Qi}\cdot u^j-u_i\cdot \,^Qs^j-g_{i\overline{k}%
}\cdot \,^QS^{jk}\text{ ,}
\end{equation}

\begin{equation}  \label{IX.3.-38}
Q^{ij}=\overline{Q}_k\,^i\cdot g^{\overline{k}j}=-\,\,\,\,u^i\cdot
p_Q^j-\,^Qs^i\cdot u^j-\,^QS^{ij}\text{ ,}
\end{equation}

\begin{equation}  \label{IX.3.-39}
Q_{ij}=g_{i\overline{k}}\cdot \overline{Q}_j\,^k=-\,\,\,\,u_i\cdot
p_{Qj}-\,^Qs_i\cdot u_j-\,^QS_{ij}\text{ ,}
\end{equation}

\[
p_{Qi}=g_{i\overline{k}}\cdot p_Q^k\text{ ,\thinspace \thinspace \thinspace
\thinspace \thinspace \thinspace \thinspace \thinspace \thinspace \thinspace
\thinspace \thinspace }^Qs_i=g_{i\overline{l}}\cdot ^Qs^l\text{ ,\thinspace
\thinspace \thinspace \thinspace \thinspace \thinspace \thinspace \thinspace
\thinspace \thinspace }^QS_{ij}=g_{i\overline{k}}\cdot ^QS^{kl}\cdot g_{%
\overline{l}j}\text{ .\thinspace \thinspace \thinspace } 
\]

On the analogy of the notions in the continuum media mechanics one can
introduce the following definitions connected with the notion momentum
density of a given energy-momentum tensor:

\textit{A.} \textit{Generalized canonical energy-momentum tensor} $\theta $

(a) Momentum density of\textit{\ }$\theta $%
\[
p_\theta =\frac 1e\cdot [g(u)](\theta )(\overline{g})=\rho _\theta \cdot
u+\,^\theta \pi \text{ .} 
\]

(b) Convective momentum density of $\theta $
................................................. $\rho _\theta \cdot u$.

\textit{B. Symmetric energy-momentum tensor of Belinfante} $_sT$

(a) Momentum density of\textit{\ }$_sT$%
\[
p_T=\frac 1e\cdot [g(u)](_sT)(\overline{g})=\rho _T\cdot u+\,^T\pi \text{ .} 
\]

(b) Convective momentum density of $_sT$
................................................ $\rho _T\cdot u$.

\textit{C. Variational energy-momentum tensor of Euler-Lagrange} $Q$

(a) Momentum density of\textit{\ } $Q$%
\[
p_Q=-\,\,\,\frac 1e\cdot [g(u)](Q)(\overline{g})=-\,\,\,\rho _Q\cdot
u-\,^Q\pi \text{ .} 
\]

(b) Convective momentum density of $Q$
................................................. $\rho _Q\cdot u$.

From the covariant Noether identities for the rest mass and the conductive
momentum densities, it follows the identity for the momentum density of the
different energy-momentum tensors 
\begin{equation}  \label{IX.3.-40}
p_\theta \equiv p_T-p_Q\text{ .}
\end{equation}

If the Euler-Lagrangian equations $\delta _vL/\delta V^A\,_B=0$ for the
non-metric tensor fields are fulfilled and $_gQ=0$, then 
\begin{equation}  \label{IX.3.-41}
p_\theta =p_T\text{ .}
\end{equation}

The representations of the energy-momentum tensors of the type 2 is
analogous to the representations of the energy-momentum tensors of the type
1 by the use of $h^u$, $h_u$ and the momentum density.

\section{Energy-momentum tensors and the energy flux density}

In the first section the notion of energy flux density has been introduced
for a given energy-momentum tensor $G$ as 
\[
e_G=G(u)=e\cdot (\rho _G\cdot u+\,^Gs)\text{ .} 
\]

The energy-momentum tensors can be now represented by the use of the
projective metrics $h^u$, $h_u$ and the energy flux density in the forms 
\[
\begin{array}{c}
G=\frac 1e\cdot e_G\otimes g(u)+u\otimes g(^G\pi )+(^GS)g= \\ 
=G_\alpha \,^\beta \cdot e_\beta \otimes e^\alpha =G_i\,^j\cdot \partial
_j\otimes dx^i\,\,\text{,}
\end{array}
\]

\begin{equation}  \label{IX.4.-1}
(G)\overline{g}=\frac 1e\cdot e_G\otimes u+u\otimes \,^G\pi +\,^GS=G^{\alpha
\beta }\cdot e_\alpha \otimes e_\beta =G^{ij}\cdot \partial _i\otimes
\partial _j\text{ ,}
\end{equation}

\begin{equation}  \label{IX.4.-2}
\begin{array}{c}
g(G)=\frac 1e\cdot g(e_G)\otimes g(u)+g(u)\otimes g(^G\pi )+g(^GS)g= \\ 
=G_{\alpha \beta }\cdot e^\alpha \otimes e^\beta =G_{ij}\cdot dx^i\otimes
dx^j\text{ .}
\end{array}
\end{equation}

The Kronecker tensor $Kr$ can also be represented by the use of $e_{Kr}$,
where 
\begin{equation}  \label{IX.4.-3}
e_{Kr}=Kr(u)=k\cdot u+e\cdot \,^{Kr}s=e\cdot (\frac 1e\cdot k\cdot
u+\,^{Kr}s)\text{ .}
\end{equation}

The structure of $e_G$ allows the introduction of the abbreviations:

(a) \textit{Convective energy flux density} of the energy-momentum

tensor $G$
............................................................................$%
\rho _G\cdot e\cdot u$.

(b) \textit{Conductive energy flux density} of the energy-momentum

tensor $G$
...............................................................................%
$e.\cdot s$.

The generalized canonical energy-momentum tensor $\theta $ can be
represented in the form 
\begin{equation}  \label{IX.4.-4}
\begin{array}{c}
\theta =\frac 1e\cdot e_\theta \otimes g(u)+u\otimes g(^\theta \pi
)+(^\theta S)g= \\ 
=\overline{\theta }_\alpha \,^\beta \cdot e_\beta \otimes e^\alpha =%
\overline{\theta }_i\,^j\cdot \partial _j\otimes dx^i\,\,\text{,}
\end{array}
\end{equation}

\noindent where 
\[
e_\theta =\theta (u)=e\cdot (\rho _\theta .u+\,^\theta s)=e_\theta ^\alpha
\cdot e_\alpha =e_\theta ^i\cdot \partial _i\text{ ,} 
\]

\noindent or by means of the forms 
\begin{equation}  \label{IX.4.-5}
(\theta )\overline{g}=\frac 1e\cdot e_\theta \otimes u+u\otimes \,^\theta
\pi +\,^\theta S=\theta ^{\alpha \beta }\cdot e_\alpha \otimes e_\beta
=\theta ^{ij}\cdot \partial _i\otimes \partial _j\text{ ,}
\end{equation}

\begin{equation}  \label{IX.4.-6}
\begin{array}{c}
g(\theta )=\frac 1e\cdot g(e_\theta )\otimes g(u)+g(u)\otimes g(^\theta \pi
)+g(^\theta S)g= \\ 
=\theta _{\alpha \beta }\cdot e^\alpha \otimes e^\beta =\theta _{ij}\cdot
dx^i\otimes dx^j\text{ .}
\end{array}
\end{equation}

In a co-ordinate basis $\theta $, $(\theta )\overline{g}$ and $g(\theta )$
will have the forms

\begin{equation}  \label{IX.4.-10}
\overline{\theta }_i\text{ }^j=\frac 1e\cdot u_i\cdot e_\theta ^j+\,^\theta
\pi _i\cdot u^j+g_{i\overline{k}}\cdot \,^\theta S^{jk}\text{ ,}
\end{equation}

\begin{equation}  \label{IX.4.-11}
\theta ^{ij}=\overline{\theta }_k\,^i\cdot g^{\overline{k}j}=\frac 1e\cdot
e_\theta ^i\cdot u^j+u^i\cdot \,^\theta \pi ^j+\,^\theta S^{ij}\text{ ,}
\end{equation}

\begin{equation}  \label{IX.4.-12}
\theta _{ij}=g_{i\overline{k}}\cdot \overline{\theta }_j\,^k=\frac 1e\cdot
e_{\theta i}\cdot u_j+u_i\cdot \,^\theta \pi _j+\,^\theta S_{ij}\text{ ,}
\end{equation}

\[
e_{\theta i}=g_{i\overline{k}}\cdot e_\theta ^k\text{ ,\thinspace \thinspace
\thinspace \thinspace \thinspace }^\theta S_{ij}=g_{i\overline{k}}\cdot
\,^\theta S^{kl}.g_{\overline{l}j}\text{ .} 
\]

The symmetric energy-momentum of Belinfante $_sT$ can be represented by the
use of $e_T$ in the form

\begin{equation}  \label{IX.4.-13}
\begin{array}{c}
_sT=\frac 1e\cdot e_T\otimes g(u)+u\otimes g(^T\pi )+(^TS)g= \\ 
=\,_sT_\alpha \,^\beta \cdot e_\beta \otimes e^\alpha =\,_sT_i\,^j\cdot
\partial _j\otimes dx^i\,\,\text{,}
\end{array}
\end{equation}

\noindent where 
\begin{equation}  \label{IX.4.-16}
e_T=\,_sT(u)=e\cdot (\rho _T\cdot u+\,^Ts)=e_T^\alpha \cdot e_\alpha
=e_T^i\cdot \partial _i\text{ ,}
\end{equation}

\noindent or by means of the forms 
\begin{equation}  \label{IX.4.-14}
(_sT)\overline{g}=\frac 1e\cdot e_T\otimes u+u\otimes \,^T\pi
+\,^TS=\,_sT^{\alpha \beta }\cdot e_\alpha \otimes e_\beta =\,_sT^{ij}\cdot
\partial _i\otimes \partial _j\text{ ,}
\end{equation}

\begin{equation}  \label{IX.4.-15}
\begin{array}{c}
g(_sT)=\frac 1e\cdot g(e_T)\otimes g(u)+g(u)\otimes g(^T\pi )+g(^TS)g= \\ 
=\,_sT_{\alpha \beta }\cdot e^\alpha \otimes e^\beta =\,_sT_{ij}\cdot
dx^i\otimes dx^j\text{ .}
\end{array}
\end{equation}

In a co-ordinate basis $_sT$, $(_sT)\overline{g}$ and $g(_sT)$ will have the
forms

\begin{equation}  \label{IX.4.-20}
_sT_i\text{ }^j=\frac 1e\cdot u_i\cdot e_T^j+\,^T\pi _i\cdot u^j+g_{i%
\overline{k}}\cdot \,^TS^{jk}\text{ ,}
\end{equation}

\begin{equation}  \label{IX.4.-21}
_sT^{ij}=\,_sT_k\,^i\cdot g^{\overline{k}j}=\frac 1e\cdot e_T^i\cdot
u^j+u^i\cdot \,^T\pi ^j+\,^TS^{ij}\text{ ,}
\end{equation}

\begin{equation}  \label{IX.4.-22}
_sT_{ij}=g_{i\overline{k}}\cdot \,_sT_j\,^k=\frac 1e\cdot e_{Ti}\cdot
u_j+u_i\cdot \,^T\pi _j+\,^TS_{ij}\text{ ,}
\end{equation}

\[
e_{Ti}=g_{i\overline{k}}\cdot e_T^k\text{ ,\thinspace \thinspace \thinspace
\thinspace \thinspace }^TS_{ij}=g_{i\overline{k}}\cdot \,^TS^{kl}\cdot g_{%
\overline{l}j}\text{ .} 
\]

The variational energy-momentum tensor of Euler-Lagrange $Q$ can be
represented in an analogous way by the use of the energy flux density $e_Q$
in the forms

\begin{equation}  \label{IX.4.-23}
\begin{array}{c}
Q=-\,\,\,\frac 1e\cdot e_Q\otimes g(u)-u\otimes g(^Q\pi )-(^QS)g= \\ 
=Q_\alpha \,^\beta \cdot e_\beta \otimes e^\alpha =Q_i\,^j\cdot \partial
_j\otimes dx^i\,\,\text{,}
\end{array}
\end{equation}

\noindent where 
\begin{equation}  \label{IX.4.-26}
e_Q=-\,\,\,Q(u)=e\cdot (\rho _Q\cdot u+\,^Qs)=e_Q^\alpha \cdot e_\alpha
=e_Q^i\cdot \partial _i\text{ ,}
\end{equation}

\noindent or by means of the forms 
\begin{equation}  \label{IX.4.-24}
(Q)\overline{g}=-\,\,\,\frac 1e\cdot e_Q\otimes u-u\otimes \,^Q\pi
-\,^QS=Q^{\alpha \beta }\cdot e_\alpha \otimes e_\beta =Q^{ij}\cdot \partial
_i\otimes \partial _j\text{ ,}
\end{equation}

\begin{equation}  \label{IX.4.-25}
\begin{array}{c}
g(Q)=-\,\,\,\frac 1e\cdot g(e_Q)\otimes g(u)-g(u)\otimes g(^Q\pi )-g(^QS)g=
\\ 
=Q_{\alpha \beta }\cdot e^\alpha \otimes e^\beta =Q_{ij}\cdot dx^i\otimes
dx^j\text{ .}
\end{array}
\end{equation}

In a co-ordinate basis $Q$, $(Q)\overline{g}$ and $g(Q)$ will have the forms

\[
\overline{Q}_i\text{ }^j=-\,\,\,\frac 1e\cdot u_i\cdot e_Q^j-\,^Q\pi _i\cdot
u^j-g_{i\overline{k}}\cdot \,^QS^{jk}\text{ ,} 
\]

\[
Q^{ij}=\overline{Q}_k\,^i\cdot g^{\overline{k}j}=-\,\,\,\frac 1e\cdot
e_Q^i\cdot u^j-u^i\cdot ^Q\pi ^j-\,^QS^{ij}\text{ ,} 
\]

\[
Q_{ij}=g_{i\overline{k}}\cdot \overline{Q}_j\,^k=-\,\,\,\frac 1e\cdot
e_{Qi}\cdot u_j-u_i\cdot \,^Q\pi _j-\,^QS_{ij}\text{ ,} 
\]

\[
e_{Qi}=g_{i\overline{k}}\cdot e_Q^k\text{ ,\thinspace \thinspace \thinspace
\thinspace \thinspace }^QS_{ij}=g_{i\overline{k}}\cdot \,^QS^{kl}\cdot g_{%
\overline{l}j}\text{ .} 
\]

On the analogy of the determined notions, the following abbreviations can be
introduced for a given energy-momentum tensor:

\textit{A. Generalized canonical energy-momentum tensor }$\theta $\textit{.}

(a) Energy flux density of $\theta $ ..................................$%
e_\theta $.

(b) Convective energy flux density of $\theta $ ...........$\rho _\theta
\cdot e\cdot u$.

\textit{B. Symmetric energy-momentum tensor of Belinfante }$_sT$\textit{.}

(a) Energy flux density of $_sT$ .................................$e_T$.

(b) Convective energy flux density of $_sT$\textit{\ ..........}$\rho
_T\cdot e\cdot u$\textit{.}

\textit{C. Variational energy-momentum tensor of Euler-Lagrange }$Q$\textit{.%
}

(a) Energy flux density of $Q$ ..................................$e_Q$.

(b) Convective energy flux density of $Q$\textit{\ ...........}$\rho _Q\cdot
e\cdot u$\textit{.}

By means of the covariant Noether identities for the rest mass density and
for the conductive energy flux density the identity for the energy flux
density follows in the form 
\begin{equation}  \label{IX.4.-31}
e_\theta \equiv e_T-e_Q\text{ .}
\end{equation}

The relation between the momentum density and the energy flux density
follows from the structure of their definitions 
\[
p_G=\rho _G\cdot u+\,^G\pi \text{ ,\thinspace \thinspace \thinspace
\thinspace \thinspace \thinspace \thinspace \thinspace \thinspace \thinspace
\thinspace }e_G=(\rho _G\cdot u+\,^Gs)\cdot e\text{ ,\thinspace \thinspace
\thinspace \thinspace \thinspace }G\sim (\theta \text{, }_sT\text{,
\thinspace }Q)\text{ .} 
\]

From the last expressions the relations 
\begin{equation}  \label{IX.4.-38}
p_G=\frac 1e\cdot e_G+\,^G\pi -\,^Gs\text{ ,}
\end{equation}

\begin{equation}  \label{IX.4.-39}
e_G=(p_G+\,^Gs-\,^G\pi )\cdot e\text{ }
\end{equation}

\noindent follow.

By the use of the different representations of the energy-momentum tensors
the different physical processes can be investigated. The application of an
representation will depend on the role of the considered quantity (rest mass
density, momentum density or energy flux density) in the dynamical process.

The physical interpretation of the introduced notions has been used for
describing Lagrangian systems in $V_4$-spaces, where the contravariant
vector field $u$ has been interpreted as a time-like vector field tangential
to the trajectories of the moving in the space-time particles. A $V_4$-space
is considered as a model of the space-time.

\section{Covariant divergency of a mixed tensor field}

The operation of the covariant differentiation along a contravariant vector
field can be extended to covariant differentiation along a contravariant
tensor field.

The Lie derivative $\pounds _\xi u$ of a contravariant vector field $u$
along a contravariant vector field $\xi $ can be expressed by the use of the
covariant differential operators $\nabla _\xi $ and $\nabla _u$ in the form 
\[
\pounds _\xi u=\nabla _\xi u-\nabla _u\xi -T(\xi ,u)\text{ ,\thinspace
\thinspace \thinspace }\xi \text{, }u\in T(M)\text{ ,} 
\]

\noindent where $T(\xi ,u)$ is the contravariant torsion vector field 
\[
T(\xi ,u)=T_{\alpha \beta }\,^\gamma \cdot \xi ^\alpha \cdot u^\beta \cdot
e_\gamma =T_{ij}\,^k\cdot \xi ^i\cdot u^j\cdot \partial _k\text{ ,} 
\]

\noindent constructed by means of the components $T_{\alpha \beta }\,^\gamma 
$ (or $T_{ij}\,^k)$ of the contravariant torsion tensor field $T$.

The Lie derivative $\pounds _\xi V$ of a contravariant tensor field $%
V=V^A\cdot e_A=V^A\cdot \partial _A\in \otimes ^m(M)$ along a contravariant
vector field $\xi $ can be written on the analogy of the relation for $%
\pounds _\xi u$ and by the use of the covariant differential operator $%
\nabla _\xi $ and an operator $\nabla _V$ in the form 
\[
\pounds _\xi V=\nabla _\xi V-\nabla _V\xi -T(\xi ,V)\text{ ,\thinspace
\thinspace \thinspace \thinspace \thinspace }\xi \in T(M)\,\,\text{,
\thinspace }V\in \otimes ^m(M)\text{ ,\thinspace \thinspace \thinspace
\thinspace } 
\]

\noindent where 
\[
\begin{array}{c}
\nabla _V\xi =-\xi ^\alpha \,_{/\beta }\cdot S_{B\alpha }\,^{A\beta }\cdot
V^B\cdot e_A=-\xi ^i\,_{;j}\cdot S_{Ci}\,^{Aj}\cdot V^C\cdot \partial _A 
\text{ ,} \\ 
T(\xi ,V)=T_{B\gamma }\,^A\cdot \xi ^\gamma \cdot V^B\cdot
e_A=T_{Ck}\,^A\cdot \xi ^k\cdot V^C\cdot \partial _A \text{ ,} \\ 
T_{B\gamma }\,^A=T_{\beta \gamma }\,^\alpha \cdot S_{B\alpha }\,^{A\beta }%
\text{ ,\thinspace \thinspace \thinspace \thinspace \thinspace \thinspace }%
T_{Ck}\,^A=T_{jk}\,^i\cdot S_{Ci}\,^{Aj}\text{ .}
\end{array}
\]

$\nabla _V\xi $ appears as a definition of the action of the operator $%
\nabla _V$ on the vector field $\xi $. Let we now consider more closely this
operator and its properties.

Let a mixed tensor field $K\in \otimes ^k\,_l(M)$ be given in a
non-co-ordinate (or co-ordinate) basis 
\[
\begin{array}{c}
K=K^C\,_D\cdot e_C\otimes e^D=K^{C_1\alpha }\,_D\cdot e_{C_1}\otimes
e_\alpha \otimes e^D \text{ ,} \\ 
e_C=e_{C_1}\otimes e_\alpha \text{ ,\thinspace \thinspace \thinspace
\thinspace \thinspace \thinspace \thinspace \thinspace \thinspace \thinspace
\thinspace \thinspace \thinspace \thinspace \thinspace \thinspace \thinspace
\thinspace \thinspace \thinspace }e^D=e^{\alpha _1}\otimes ...\otimes
e^{\alpha _l}\text{ .}
\end{array}
\]

The action of the operator $\nabla _V$ on the mixed tensor field $K$ can be
defined on the analogy of the action of $\nabla _V$ on a contravariant
vector field $\xi $%
\begin{equation}  \label{X.1.-1}
\nabla _VK=-K^{C_1\alpha }\,_{D/\beta }\cdot S_{B\alpha }\,^{A\beta }\cdot
V^B\cdot e_{C_1}\otimes e^D\otimes e_A\text{ ,}
\end{equation}

\noindent where $\nabla _V$ is the covariant differential operator along a
contravariant tensor field $V$%
\[
\nabla _V:K\Rightarrow \nabla _VK\text{ , \thinspace \thinspace \thinspace
\thinspace \thinspace \thinspace \thinspace \thinspace }K\in \otimes
^k\,_l(M)\text{ ,\thinspace \thinspace \thinspace \thinspace \thinspace
\thinspace }V\in \otimes ^m(M)\text{ ,\thinspace \thinspace \thinspace
\thinspace }\nabla _VK\in \otimes ^{k-1+m}\,_l(M)\text{ .} 
\]

\textit{Remark}. There is an other possibility for a generalization of the
action of $\nabla _V$ on a mixed tensor field 
\[
\begin{array}{c}
\nabla _VK=-\sum_{m=1}^kK^{\alpha _1...\alpha _m...\alpha _k}\,_{D/\beta
}.S_{B\alpha _m}\,^{A\beta }.V^B.e_{\alpha _1}\otimes ...\otimes e_{\alpha
_{m-1}}\otimes e_{\alpha _{m+1}}\otimes \\ 
...\otimes e_{\alpha _k}\otimes e^D\otimes e_A\text{ .}
\end{array}
\]

The result of the action of this operator on a contravariant vector field $%
\xi $ is identical with the action of the above defined operator $\nabla _V$.

\textit{Remark}. A covariant differential operator along a contravariant
tensor field can also be defined through its action on mixed tensor fields
in the form 
\[
\overline{\nabla }_VK=K^C\,_{D/\beta }.V^{A_1\beta }.e_C\otimes e^D\otimes
e_{A_1}\text{ .} 
\]

The operator $\overline{\nabla }_V$ differs from $\nabla _V$ in its action
on a contravariant vector field and does not appear as a generalization of
the already defined operator by its action on a contravariant vector field.
It appears as a new differential operator acting on mixed tensor fields.

The \textit{covariant differential operator} $\nabla _V$ has the properties:

(a) Linear operator 
\[
\begin{array}{c}
\nabla _V(\alpha \cdot K_1+\beta \cdot K_2)=\alpha \cdot \nabla _VK_1+\beta
\cdot \nabla _VK_2 \text{ ,} \\ 
\alpha \text{ , }\beta \in R\text{ (or }C\text{), \thinspace \thinspace
\thinspace \thinspace \thinspace \thinspace \thinspace \thinspace \thinspace
\thinspace \thinspace \thinspace \thinspace }K_1,K_2\in \otimes ^k\,_l(M)%
\text{ .}
\end{array}
\]

The proof of this property follows immediately from (\ref{X.1.-1}) and the
linear property of the covariant differential operator along a basic
contravariant vector field.

(b) Differential operator (not obeying the Leibniz rule) 
\[
\begin{array}{c}
\nabla _V(K\otimes S)=\nabla _{e_\beta }K\otimes \overline{S}\,^\beta
+K\otimes \nabla _VS\text{ ,} \\ 
K=K^A\,_B\cdot e_A\otimes e^B \text{ ,\thinspace \thinspace \thinspace
\thinspace \thinspace \thinspace }\nabla _{e_\beta }K=K^A\,_{B/\beta }\cdot
e_A\otimes e^B\text{ ,} \\ 
S= \widetilde{S}\,^C\,_D\cdot e_C\otimes e^D=\widetilde{S}\,^{C_1\alpha
}\,_D\cdot e_{C_1}\otimes e_\alpha \otimes e^D\text{ ,} \\ 
\overline{S}\,^\beta =-\widetilde{S}\,^{C_1\alpha }\,_D\cdot S_{E\alpha
}\,^{F\beta }\cdot V^E\cdot e_{C_1}\otimes e^D\otimes e_F\text{ .}
\end{array}
\]

The proof of this property follows from the action of the defined in (\ref
{X.1.-1}) operator $\nabla _V$ and the properties of the covariant
derivative of the product of the components of the tensor fields $K$ and $S$.

If the tensor field $V$ is given as a contravariant metric tensor field $%
\overline{g}$, then the covariant differential operator $\nabla _V$ ($V=%
\overline{g}$) will have additional properties connected with the properties
of the contravariant metric tensor field.

%TCIMACRO{
%\TeXButton{definition}{\begin{definition}
%{\it Contravariant metric differential operator} $\nabla _{\overline{g}}$.
%Covariant differential operator $\nabla _V$ for $V=\overline{g}$.
%\end{definition}
%}}
%BeginExpansion
\begin{definition}
{\it Contravariant metric differential operator} $\nabla _{\overline{g}}$.
Covariant differential operator $\nabla _V$ for $V=\overline{g}$.
\end{definition}
%
%EndExpansion

By means of the relations 
\begin{equation}  \label{X.1.-2}
\begin{array}{c}
-S_{B\alpha }\,^{A\beta }\cdot g^B\cdot e_A=(g_\alpha ^\sigma \cdot g^{\beta
\kappa }+g_\alpha ^\kappa \cdot g^{\beta \sigma }).e_\sigma \otimes e_\kappa
= \\ 
=(g_\alpha ^\sigma \cdot g^{\beta \kappa }+g_\alpha ^\kappa \cdot g^{\beta
\sigma }).e_\sigma .e_\kappa \text{ ,}
\end{array}
\end{equation}
\[
e_\sigma .e_\kappa =\frac 12(e_\sigma \otimes e_\kappa +e_\kappa \otimes
e_\sigma )\text{ ,} 
\]
\begin{equation}  \label{X.1.-3}
K^{C_1\alpha }\,_{D/\beta }\cdot g_\alpha ^\sigma =K^{C_1\sigma }\,_{D/\beta
}\text{ ,}
\end{equation}

\noindent the action of the contravariant metric differential operator on a
mixed tensor field $K$ can be represented in the form 
\begin{equation}  \label{X.1.-4}
\begin{array}{c}
\nabla _{\overline{g}}K=(K^{C_1\sigma }\,_{D/\beta }\cdot g^{\beta \kappa
}+K^{C_1\kappa }\,_{D/\beta }\cdot g^{\beta \sigma })\cdot e_{C_1}\otimes
e^D\otimes e_\sigma \otimes e_\kappa = \\ 
(K^{C_1\sigma }\,_{D/\beta }\cdot g^{\beta \kappa }+K^{C_1\kappa
}\,_{D/\beta }\cdot g^{\beta \sigma }).e_{C_1}\otimes e^D\otimes e_\sigma
.e_\kappa \text{ .}
\end{array}
\end{equation}

The properties of the operator $\nabla _{\overline{g}}$ are determined
additionally by the properties of the contravariant metric tensor field of
second rank:

(a) $\nabla _{\overline{g}}:K\Rightarrow \nabla _{\overline{g}}K$
,\thinspace \thinspace \thinspace \thinspace $K\in \otimes ^k\,_l(M)$, $%
\nabla _{\overline{g}}K\in \otimes ^{k+1}\,_l(M)$.

(b) Linear operator 
\[
\nabla _{\overline{g}}(\alpha \cdot K_1+\beta \cdot K_2)=\alpha \cdot \nabla
_{\overline{g}}K_1+\beta \cdot \nabla _{\overline{g}}K_2\text{ . } 
\]

(c) Differential operator (not obeying the Leibniz rule) 
\begin{equation}  \label{X.1.-5}
\begin{array}{c}
\nabla _{\overline{g}}(K\otimes S)=\nabla _{e_\beta }K\otimes \overline{S}%
\,^\beta +K\otimes \nabla _{\overline{g}}S\text{ ,} \\ 
K\in \otimes ^k\,_l(M) \text{ ,\thinspace \thinspace \thinspace \thinspace
\thinspace \thinspace \thinspace \thinspace \thinspace \thinspace \thinspace 
}S\in \otimes ^m\,_r(M)\text{ ,} \\ 
K=K^A\,_B\cdot e_A\otimes e^B \text{ ,\thinspace \thinspace \thinspace
\thinspace \thinspace \thinspace }\nabla _{e_\beta }K=K^A\,_{B/\beta }\cdot
e_A\otimes e^B\text{ ,} \\ 
S= \widetilde{S}\,^C\,_D\cdot e_C\otimes e^D=\widetilde{S}\,^{C_1\alpha
}\,_D\cdot e_{C_1}\otimes e_\alpha \otimes e^D\text{ ,} \\ 
\overline{S}\,^\beta =(\widetilde{S}\,^{C_1\sigma }\,_D\cdot g^{\beta \kappa
}+\widetilde{S}\,^{C_1\kappa }\,_D\cdot g^{\beta \sigma })\cdot
e_{C_1}\otimes e^D\otimes e_\sigma .e_\kappa \text{ .}
\end{array}
\end{equation}

\textit{Remark}. 
%TCIMACRO{
%\TeXButton{remark}{The definition of $\nabla _{\overline{g}}$ in (\ref{X.1.-4}) differs from
%the definition in \cite{Manoff-2}, where $\nabla _{\overline{g}}\equiv 
%\overline{\nabla }_{\overline{g}}$, i. e. the contravariant metric
%differential operator is defined in the last case as a special case of the
%covariant differential operator $\overline{\nabla }_V$ for $V=\overline{g}$.
%
%}}
%BeginExpansion
The definition of $\nabla _{\overline{g}}$ in (\ref{X.1.-4}) differs from
the definition in \cite{Manoff-2}, where $\nabla _{\overline{g}}\equiv 
\overline{\nabla }_{\overline{g}}$, i. e. the contravariant metric
differential operator is defined in the last case as a special case of the
covariant differential operator $\overline{\nabla }_V$ for $V=\overline{g}$.

%
%EndExpansion

The notion of covariant divergency of a mixed tensor field has been used in $%
V_4$-spaces for the determination of conditions for the existence of local
conserved quantities and in identities of the type of the first covariant
Noether identity. Usually, the covariant divergency of a contravariant or
mixed tensor field has been given in co-ordinate or non-co-ordinate basis in
the form 
\begin{equation}  \label{X.1.-6}
\delta K=K^{A\beta }\,_{B/\beta }\cdot e_A\otimes e^B=K^{Ci}\,_{D;i}\cdot
\partial _C\otimes dx^D\text{ ,}
\end{equation}

\noindent where 
\begin{equation}  \label{X.1.-7}
K^{A\beta }\,_{B/\beta }=K^{A\beta }\,_{B/\gamma }\cdot g_\beta ^\gamma 
\text{ ,\thinspace \thinspace \thinspace \thinspace \thinspace \thinspace
\thinspace \thinspace \thinspace }K^{Ci}\,_{D;i}=K^{Ci}\,_{D;k}\cdot g_i^k%
\text{ .\thinspace }
\end{equation}

For full anti-symmetric covariant tensor fields (differential forms) the
covariant divergency (called also codifferential) $\delta $ is defined by
means of the Hodge operator $*$, its reverse operator $*^{-1}$ and the
external differential operator $_a\overline{d}$ in the form \cite{Dubrovin}, 
\cite{von Westenholz} (pp. 147-149) 
\begin{equation}  \label{X.1.-8}
\delta =*^{-1}\circ \,_a\overline{d}\circ *\text{ .}
\end{equation}

\textit{Remark}. 
%TCIMACRO{
%\TeXButton{remark}{The {\it Hodge operator} is constructed by means of the permutation
%(Levi-Chivita) symbols. It maps a full covariant anti-symmetric tensor of
%rank $(0,p)\equiv \,^a\otimes _pM)\equiv \Lambda ^p(M)$ in a full covariant
%anti-symmetric tensor of rank $(0,n-p)\equiv \,^a\otimes _{n\,-\,p}(M)\equiv
%\Lambda ^{n\,-\,p}(M)$, where $\dim M=n$, 
%\[
%\ast \,:\,_aA\rightarrow *\,_aA\text{ , \thinspace \thinspace \thinspace
%\thinspace \thinspace \thinspace \thinspace \thinspace }_aA\in \Lambda ^p(M)\text{ , \thinspace \thinspace \thinspace \thinspace \thinspace \thinspace
%\thinspace }*\,_aA\in \Lambda ^{n\,-\,p}(M)\text{ ,} 
%\]
%
%with 
%\[
%_aA=A_{[i_1...i_p]}.dx^{i_1}\wedge ...\wedge dx^{i_p}\text{ ,\thinspace
%\thinspace \thinspace \thinspace \thinspace \thinspace \thinspace \thinspace
%\thinspace \thinspace }*\,_aA=*A_{[j_1...j_{n\,-\,p}]}.dx^{j_1}\wedge
%...\wedge dx^{j_{n\,-\,p}}\text{ ,} 
%\]
%\[
%\ast A_{[j_1...j_{n\,-\,p}]}=\frac 1{p!}.\varepsilon
%_{i_1...i_pj_1...j_{n\,-\,p}}.A^{[i_1...i_p]}\text{ , \thinspace \thinspace }A^{[i_1...i_p]}=g^{i_1\overline{k}_1}....g^{i_p\overline{k}_p}.A_{[k_1...k_p]}\text{ ,\thinspace \thinspace \thinspace \thinspace } 
%\]
%\[
%\ast ^{-1}=(-1)^{p.(n-p)}.*\text{ .} 
%\]
%
%}}
%BeginExpansion
The {\it Hodge operator} is constructed by means of the permutation
(Levi-Chivita) symbols. It maps a full covariant anti-symmetric tensor of
rank $(0,p)\equiv \,^a\otimes _pM)\equiv \Lambda ^p(M)$ in a full covariant
anti-symmetric tensor of rank $(0,n-p)\equiv \,^a\otimes _{n\,-\,p}(M)\equiv
\Lambda ^{n\,-\,p}(M)$, where $\dim M=n$, 
\[
\ast \,:\,_aA\rightarrow *\,_aA\text{ , \thinspace \thinspace \thinspace
\thinspace \thinspace \thinspace \thinspace \thinspace }_aA\in \Lambda ^p(M)\text{ , \thinspace \thinspace \thinspace \thinspace \thinspace \thinspace
\thinspace }*\,_aA\in \Lambda ^{n\,-\,p}(M)\text{ ,} 
\]

with 
\[
_aA=A_{[i_1...i_p]}.dx^{i_1}\wedge ...\wedge dx^{i_p}\text{ ,\thinspace
\thinspace \thinspace \thinspace \thinspace \thinspace \thinspace \thinspace
\thinspace \thinspace }*\,_aA=*A_{[j_1...j_{n\,-\,p}]}.dx^{j_1}\wedge
...\wedge dx^{j_{n\,-\,p}}\text{ ,} 
\]
\[
\ast A_{[j_1...j_{n\,-\,p}]}=\frac 1{p!}.\varepsilon
_{i_1...i_pj_1...j_{n\,-\,p}}.A^{[i_1...i_p]}\text{ , \thinspace \thinspace }A^{[i_1...i_p]}=g^{i_1\overline{k}_1}....g^{i_p\overline{k}_p}.A_{[k_1...k_p]}\text{ ,\thinspace \thinspace \thinspace \thinspace } 
\]
\[
\ast ^{-1}=(-1)^{p.(n-p)}.*\text{ .} 
\]

%
%EndExpansion

By the use of the contravariant metric differential operator $\nabla _{%
\overline{g}}$, the covariant metric tensor field $g$ and the contraction
operator one can introduce the notion of covariant divergency of a mixed
tensor field $K$ with finite rank.

%TCIMACRO{
%\TeXButton{definition}{\begin{definition}
%{\bf \ } {\it Covariant divergency }$\delta K${\it \ of a mixed tensor field}
%$K$\[
%\delta K=\frac 12.[\nabla _{\overline{g}}K]g=K^{A\beta }\,_{B/\beta
%}.e_A\otimes e^B=K^{Ci}\,_{D;i}.\partial _C\otimes dx^D\text{ ,} 
%\]
%\end{definition}
%}}
%BeginExpansion
\begin{definition}
{\bf \ } {\it Covariant divergency }$\delta K${\it \ of a mixed tensor field}
$K$\[
\delta K=\frac 12.[\nabla _{\overline{g}}K]g=K^{A\beta }\,_{B/\beta
}.e_A\otimes e^B=K^{Ci}\,_{D;i}.\partial _C\otimes dx^D\text{ ,} 
\]
\end{definition}
%
%EndExpansion

\noindent where 
\[
\begin{array}{c}
K=K^{A\beta }\,_B\cdot e_A\otimes e_\beta \otimes e^B=K^{Ci}\,_D\cdot
\partial _C\otimes \partial _i\otimes dx^D \text{ ,} \\ 
K\in \otimes ^k\,_l(M)\text{ , \thinspace \thinspace \thinspace \thinspace
\thinspace }k\geq 1\text{ .}
\end{array}
\]

$\delta $ is called operator of the covariant divergency 
\[
\delta :K\Rightarrow \delta K\text{ ,\thinspace \thinspace \thinspace
\thinspace \thinspace \thinspace \thinspace \thinspace \thinspace }K\in
\otimes ^k\,_l(M)\text{ ,\thinspace \thinspace \thinspace \thinspace
\thinspace \thinspace \thinspace \thinspace \thinspace \thinspace }\delta
K\in \otimes ^{k-1}\,_l(M)\text{ ,\thinspace \thinspace \thinspace
\thinspace }k\geq 1\text{ .} 
\]

\textit{Remark}. 
%TCIMACRO{
%\TeXButton{remark}{The symbol $\delta $ has also been introduced for the variation operator.
%Both operators are different from each other and can easily be
%distinguished. Ambiguity would occur only if the symbol $\delta $ is used
%out of the context. In such a case, the definition of the symbol $\delta $
%is necessary.
%}}
%BeginExpansion
The symbol $\delta $ has also been introduced for the variation operator.
Both operators are different from each other and can easily be
distinguished. Ambiguity would occur only if the symbol $\delta $ is used
out of the context. In such a case, the definition of the symbol $\delta $
is necessary.
%
%EndExpansion

The properties of the operator of the covariant divergency $\delta $ are
determined by the properties of the contravariant metric differential
operator, the contraction operator and the metric tensor fields $g$ and $%
\overline{g}$

(a) The operator of the covariant divergency $\delta $ is a linear operator 
\begin{equation}  \label{X.1.-9}
\begin{array}{c}
\delta (\alpha \cdot K_1+\beta \cdot K_2)=\alpha \cdot \delta K_1+\beta
\cdot \delta K_2 \text{ ,\thinspace \thinspace } \\ 
\text{\thinspace }\alpha ,\beta \in R\text{ (or }C\text{) ,\thinspace }%
K_1,K_2\in \otimes ^k\,_l(M)\text{ .}
\end{array}
\end{equation}

The proof of this property follows immediately from the definition of the
covariant divergency.

(b) Action on a tensor product of tensor fields 
\begin{equation}  \label{X.1.-10}
\delta (K\otimes S)=\overline{\nabla }_SK+K\otimes \delta S\text{ ,}
\end{equation}

\noindent where 
\begin{equation}  \label{X.1.-11}
\begin{array}{c}
K=K^A\,_B\cdot e_A\otimes e^B \text{ ,\thinspace \thinspace \thinspace
\thinspace \thinspace \thinspace }S=S^{C\beta }\,_D\cdot e_C\otimes e_\beta
\otimes e^D\text{ ,} \\ 
\overline{\nabla }_SK=K^A\,_{B/\beta }\cdot S^{C\beta }\,_D\cdot e_A\otimes
e^B\otimes e_C\otimes e^D\text{ \thinspace \thinspace \thinspace \thinspace
\thinspace \thinspace \thinspace (see above }\overline{\nabla }_V\text{) .}
\end{array}
\end{equation}

The proof of this property follows from the properties of $\nabla _{%
\overline{g}}$ and from the relations 
\begin{equation}  \label{X.1.-12}
\frac 12[\nabla _{e_\beta }K\otimes \overline{S}\,^\beta ]g=K^A\,_{B/\beta
}\cdot S^{C\beta }\,_D\cdot e_A\otimes e^B\otimes e_C\otimes e^D\text{ ,}
\end{equation}
\begin{equation}  \label{X.1.-13}
\frac 12[K\otimes \nabla _{\overline{g}}S]g=K\otimes \frac 12\cdot [\nabla _{%
\overline{g}}S]g=K\otimes \delta S\text{ .}
\end{equation}

(c) Action on a contravariant vector field $u$%
\begin{equation}  \label{X.1.-14}
\delta u=\frac 12\cdot [\nabla _{\overline{g}}u]g=u^\beta \,_{\,/\beta
}=u^i\,_{;i}\text{ .}
\end{equation}

(d) Action on the tensor product of two contravariant vector fields $u$ and $%
v$%
\begin{equation}  \label{X.1.-15}
\delta (u\otimes v)=\nabla _vu+\delta u\cdot v\text{ ,\thinspace \thinspace
\thinspace \thinspace \thinspace \thinspace \thinspace \thinspace }\overline{%
\nabla }_vu=\nabla _vu\text{ .}
\end{equation}

(e) Action of the product of an invariant function $L$ and a mixed tensor
field $K$%
\begin{equation}  \label{X.1.-16}
\delta (L.K)=\overline{\nabla }_KL+L\cdot \delta K\text{ ,}
\end{equation}
\begin{equation}  \label{X.1.-17}
\begin{array}{c}
\overline{\nabla }_KL=L_{/\beta }\cdot K^{A\beta }\,_B\cdot e_A\otimes e^B%
\text{ ,\thinspace \thinspace \thinspace \thinspace \thinspace \thinspace }%
\delta K=K^{A\beta }\,_{B/\beta }\cdot e_A\otimes e^B\text{ ,} \\ 
L_{/\beta }=e_\beta L \text{ ,\thinspace \thinspace \thinspace \thinspace
\thinspace }L_{;i}=L_{,i}\text{ ,} \\ 
K=K^{A\beta }\,_B\cdot e_A\otimes e_\beta \otimes e^B\in \otimes ^k\,_l(M)%
\text{ .}
\end{array}
\end{equation}

\textit{Special case}: Action of the product of an invariant function $L$
and the contravariant metric tensor $\overline{g}$: 
\begin{equation}  \label{X.1.-18}
\delta (L.\overline{g})=(L_{/\beta }\cdot g^{\alpha \beta }+L\cdot g^{\alpha
\beta }\,_{/\beta })=(L_{,j}\cdot g^{ij}+L\cdot g^{ij}\,_{;j})\text{ .}
\end{equation}

(f) Action on an anti-symmetric tensor product of two contravariant vector
fields $u$ and $v$%
\begin{equation}  \label{X.1.-19}
\begin{array}{c}
\delta (u\wedge v)=\frac 12\cdot (\nabla _vu-\nabla _uv+\delta v\cdot
u-\delta u\cdot v)= \\ 
=-\frac 12\cdot [\pounds _uv+T(u,v)+\delta u\cdot v-\delta v\cdot u]\text{ .}
\end{array}
\end{equation}

(g) Action on a full anti-symmetric contravariant tensor field $A$ of second
rank 
\begin{equation}  \label{X.1.-20}
\begin{array}{c}
\delta A=\frac 12\cdot (A^{\alpha \beta }-A^{\beta \alpha })_{/\beta }\cdot
e_\alpha =\frac 12\cdot (A^{ij}-A^{ji})_{;j}\cdot \partial _i \text{ ,} \\ 
A=A^{\alpha \beta }\cdot e_\alpha \wedge e_\beta =A^{ij}\cdot \partial
_i\wedge \partial _j\text{ ,\thinspace \thinspace \thinspace \thinspace
\thinspace \thinspace }A^{\alpha \beta }=-\,A^{\beta \alpha }\text{ .}
\end{array}
\end{equation}

(h) Action on a tensor product of a contravariant vector field $u$,
multiplied with an invariant function, and a covariant vector field $g(v)$
with the contravariant vector field $v$%
\begin{equation}  \label{X.1.-21}
\begin{array}{c}
\delta (\varepsilon \cdot u\otimes g(v))=(u\varepsilon )\cdot
g(v)+\varepsilon \cdot [\delta u\cdot g(v)+(\nabla _ug)(v)+g(\nabla _uv)]=
\\ 
=[u\varepsilon +\varepsilon \cdot \delta u]\cdot g(v)+\varepsilon \cdot
[(\nabla _ug)(v)+g(\nabla _uv)] \text{ ,} \\ 
\varepsilon \in C^r(M)\text{ , }\,\,\,\,\,\,\,\,\varepsilon ^{\prime
}(x^{k^{\prime }})=\varepsilon (x^k)\text{ ,\thinspace \thinspace \thinspace
\thinspace \thinspace \thinspace \thinspace \thinspace }u,v\in T(M)\text{ .}
\end{array}
\end{equation}

\textit{Special case}: $v\equiv u$: 
\begin{equation}  \label{X.1.-22}
\delta (\varepsilon \cdot u\otimes g(u))=[u\varepsilon +\varepsilon \cdot
\delta u]\cdot g(u)+\varepsilon \cdot [(\nabla _ug)(u)+g(a)]\text{
,\thinspace \thinspace \thinspace \thinspace }\nabla _uu=a\text{ .}
\end{equation}

\textit{Special case}: $\varepsilon =1$: 
\begin{equation}  \label{X.1.-23}
\delta (u\otimes g(v))=\delta u\cdot g(v)+(\nabla _ug)(v)+g(\nabla _uv)\text{
.}
\end{equation}

\subsection{Covariant divergency of a mixed tensor field of second rank}

From the definition of the covariant divergency $\delta K$ of a mixed tensor
field $K$, the explicit form of the covariant divergency of tensor fields of
second rank of the type 1. or 2. follows as 
\begin{equation}  \label{X.2.-1}
\delta G=\frac 12[\nabla _{\overline{g}}G]g=G_\alpha \,^\beta \,_{/\beta
}\cdot e^\alpha =G_i\,^j\,_{;j}\cdot dx^i\text{ ,}
\end{equation}
\begin{equation}  \label{X.2.-2}
\delta \overline{G}=\frac 12[\nabla _{\overline{g}}\overline{G}]g=\overline{G%
}\,^\beta \,_{\alpha /\beta }\cdot e^\alpha =\overline{G}\,^j\,_{i;j}\cdot
dx^i\text{ .}
\end{equation}

By the use of the relations (\ref{X.1.-14}) $\div \,$(\ref{X.1.-19}), (\ref
{X.1.-21}) $\div \,$(\ref{X.1.-23}), and the expression [see (\ref{X.1.-10}) 
$\div $ (\ref{X.1.-13})] 
\begin{equation}  \label{X.2.-3}
\overline{\nabla }_v(g(u))=\nabla _v(g(u))=(\nabla _vg)(u)+g(\nabla _vu)%
\text{ ,}
\end{equation}
\begin{equation}  \label{X.2.-4}
\delta (g(u)\otimes v)=\delta v\cdot g(u)+(\nabla _vg)(u)+g(\nabla _vu)\text{
,}
\end{equation}
\begin{equation}  \label{X.2.-5}
\delta ((^GS)g)=(g_{\alpha \overline{\gamma }}\cdot \,^GS^{\beta \gamma
})_{/\beta }\cdot e^\alpha \text{ ,}
\end{equation}

\noindent the covariant divergency of the representation of $G$ by means the
rest mass density $\rho _G$ ($\varepsilon _G=\rho _G$) 
\[
G=\rho _G\cdot u\otimes g(u)+u\otimes g(^G\pi )+\,^Gs\otimes g(u)+(^GS)g, 
\]

\noindent can be found in the form ($\nabla _uu=a$) 
\begin{equation}  \label{X.2.-6}
\begin{array}{c}
\delta G=\rho _G\cdot g(a)+(u\rho _G+\rho _G\cdot \delta u+\delta ^Gs)\cdot
g(u)+\delta u\cdot g(^G\pi )+g(\nabla _u\,^G\pi )+ \\ 
+\,\,\,g(\nabla _{^Gs}u)+\rho _G\cdot (\nabla _ug)(u)+(\nabla _ug)(^G\pi
)+(\nabla _{^Gs}g)(u)+\delta ((^GS)g)\text{ .}
\end{array}
\end{equation}

$\overline{g}(\delta G)$ will have the form 
\begin{equation}  \label{X.2.-7}
\begin{array}{c}
\overline{g}(\delta G)=\rho _G\cdot a+(u\rho _G+\rho _G\cdot \delta u+\delta
^Gs)\cdot u+\delta u\cdot \,^G\pi +\nabla _u\,^G\pi + \\ 
+\,\,\,\nabla _{^Gs}u+\rho _G\cdot \overline{g}(\nabla _ug)(u)+\overline{g}%
(\nabla _ug)(^G\pi )+\overline{g}(\nabla _{^Gs}g)(u)+\overline{g}(\delta
((^GS)g))\text{ .}
\end{array}
\end{equation}

In a co-ordinate basis $\delta G$ and $\overline{g}(\delta G)$ will have the
forms 
\begin{equation}  \label{X.2.-10}
\begin{array}{c}
G_i\,^j\,_{;j}=\rho _G\cdot a_i+(\rho _{G,j}\cdot u^j+\rho _G\cdot
u^j\,_{;j}+\,^Gs^j \text{ }_{;j})\cdot u_i+u^j\,_{;j}\cdot \,^G\pi _i+ \\ 
+\;g_{i \overline{j}}\cdot (^G\pi ^j\,_{;k}\cdot u^k+u^j\,_{;k}\cdot
\,^Gs^k)+g_{ij;k}\cdot (\rho _G\cdot u^k\cdot u^{\overline{j}}+u^k\cdot
\,^G\pi ^{\overline{j}}+\,^Gs^k\cdot u^{\overline{j}})+ \\ 
+\,\,(g_{i\overline{k}}\cdot ^GS^{jk})_{;j}\text{ ,}
\end{array}
\end{equation}
\[
a_i=g_{i\overline{j}}\cdot a^j\text{ ,\thinspace \thinspace \thinspace
\thinspace \thinspace }a^i=u^i\,_{;j}\cdot u^j\text{ ,\thinspace \thinspace
\thinspace \thinspace }\rho _{G;i}=\rho _{G,i}\text{ ,\thinspace \thinspace
\thinspace \thinspace \thinspace \thinspace }u_i=g_{i\overline{k}}\cdot u^k%
\text{ ,\thinspace \thinspace \thinspace \thinspace }^G\pi _i=g_{i\overline{j%
}}\cdot ^G\pi ^j\text{ ,} 
\]
\begin{equation}  \label{X.2.-11}
\begin{array}{c}
g^{i \overline{k}}\cdot G_k\,^j\,_{;j}=\rho _G\cdot a^i+(\rho _{G,j}\cdot
u^j+\rho _G\cdot u^j\,_{;j}+\,^Gs^j\text{ }_{;j})\cdot u^i+u^j\,_{;j}\cdot
\,^G\pi ^i+ \\ 
+\;^G\pi ^i\,_{;j}\cdot u^j+u^i\,_{;j}\cdot \,^Gs^j+g^{i \overline{l}}\cdot
g_{lj;k}\cdot (\rho _G\cdot u^k\cdot u^{\overline{j}}+u^k\cdot \,^G\pi ^{%
\overline{j}}+\,^Gs^k\cdot u^{\overline{j}})+ \\ 
+\,\,g^{i\overline{l}}\cdot (g_{lk}\cdot \,^GS^{jk})_{;j}\text{ .}
\end{array}
\end{equation}

The relation between $\delta G$ and $\delta \overline{G}$ follows from the
relation between $G$ and $\overline{G}$%
\begin{equation}  \label{X.2.-18}
\overline{G}=g(G)\overline{g}:\delta \overline{G}=\delta (g(G)\overline{g}%
)=\frac 12[\nabla _{\overline{g}}(g(G)\overline{g})]\text{ .}
\end{equation}

The \textit{covariant divergency of the Kronecker tensor field} can be found
in an analogous way as the covariant divergency of a tensor field of second
rank of the type 1, since $Kr=g_\beta ^\alpha \cdot e_\alpha \otimes e^\beta
=g_j^i\cdot \partial _i\otimes dx^j$%
\begin{equation}  \label{X.2.-37}
\delta Kr=\frac 12[\nabla _{\overline{g}}Kr]g=g_\alpha ^\beta \,_{/\beta
}\cdot e^\alpha =g_{i;j}^j\cdot dx^i\text{ .}
\end{equation}

If we use the representation of $Kr$%
\[
Kr=\frac 1e\cdot k\cdot u\otimes g(u)+u\otimes g(^{Kr}\pi )+\,^{Kr}s\otimes
g(u)+(^{Kr}S)g\text{ ,} 
\]

\noindent then the covariant divergency $\delta Kr$ can be written in the
form 
\begin{equation}  \label{X.2.-38}
\begin{array}{c}
\delta Kr=\frac 1e\cdot k\cdot g(a)+[u(\frac 1e\cdot k)+\frac 1e\cdot k\cdot
\delta u+\delta ^{Kr}s]\cdot g(u)+ \\ 
+\delta u\cdot g(^{Kr}\pi )+g(\nabla _u\,^{Kr}\pi )+g(\nabla _{^{Kr}s}u)+ \\ 
+\,\frac 1e\cdot k\cdot (\nabla _ug)(u)+(\nabla _ug)(^{Kr}\pi )+(\nabla
_{^{Kr}s}g)(u)+\delta ((^{Kr}S)g)\text{ ,}
\end{array}
\end{equation}

\noindent or in the form 
\begin{equation}  \label{X.2.-39}
\begin{array}{c}
\overline{g}(\delta Kr)=\frac 1e\cdot k\cdot a+[u(\frac 1e\cdot k)+\frac
1e\cdot k\cdot \delta u+\delta ^{Kr}s]\cdot u+ \\ 
+\,\delta u\cdot \,^{Kr}\pi +\nabla _u\,^{Kr}\pi +\nabla _{^{Kr}s}u+ \\ 
+\,\frac 1e\cdot k\cdot \overline{g}(\nabla _ug)(u)+\overline{g}(\nabla
_ug)(^{Kr}\pi )+\overline{g}(\nabla _{^{Kr}s}g)(u)+\overline{g}(\delta
((^{Kr}S)g))\text{ .}
\end{array}
\end{equation}

In a co-ordinate basis $\delta Kr$ and $\overline{g}(\delta Kr)$ will have
the forms 
\begin{equation}  \label{X.2.-42}
\begin{array}{c}
g_i^j\,_{;j}=\frac 1e\cdot k\cdot a_i+[(\frac 1e\cdot k)_{,j}\cdot u^j+\frac
1e\cdot k\cdot u^j\,_{;j}+\,^{Kr}s^j\,_{;j}]\cdot u_i+ \\ 
+\,\,u^j\,_{;j}\cdot \,^{Kr}\pi _i+g_{i \overline{j}}\cdot (^{Kr}\pi
^j\,_{;k}\cdot u^k+u^j\,_{;k}\cdot \,^{Kr}s^k)+ \\ 
+\,g_{ij;k}\cdot (\frac 1e\cdot k\cdot u^{\overline{j}}\cdot u^k+\,^{Kr}\pi
^{\overline{j}}\cdot u^k+u^{\overline{j}}\cdot \,^{Kr}s^k)+(g_{i\overline{k}%
}\cdot \,^{Kr}S^{jk})_{;j}\text{ ,}
\end{array}
\end{equation}
\begin{equation}  \label{X.2.-43}
\begin{array}{c}
g^{i \overline{k}}\cdot g_k^j\,_{;j}=\frac 1e\cdot k\cdot a^i+[(\frac
1e\cdot k)_{,j}\cdot u^j+\frac 1e\cdot k\cdot
u^j\,_{;j}+\,^{Kr}s^j\,_{;j}]\cdot u^i+ \\ 
+\,\,u^j\,_{;j}\cdot \,^{Kr}\pi ^i+\,^{Kr}\pi ^i\,_{;j}\cdot
u^j+u^i\,_{;j}\cdot \,^{Kr}s^j+ \\ 
+\,g^{i\overline{l}}\cdot g_{lj;k}\cdot (\frac 1e\cdot k\cdot u^{\overline{j}%
}\cdot u^k+\,^{Kr}\pi ^{\overline{j}}\cdot u^k+u^{\overline{j}}\cdot
\,^{Kr}s^k)+g^{i\overline{l}}\cdot (g_{l\overline{k}}\cdot
\,^{Kr}S^{jk})_{;j}\text{ .}
\end{array}
\end{equation}

\section{Covariant divergency of the energy-momentum tensors and the rest
mass density}

The covariant divergency of the energy-momentum tensors can be represented
by the use of the projective metrics $h^u$, $h_u$ of the contravariant
vector field $u$ and the rest mass density for the corresponding
energy-momentum tensor. In this case the representation of the
energy-momentum tensor is in the form 
\[
G=(\rho _G+\frac 1e\cdot L\cdot k)\cdot u\otimes g(u)-L\cdot Kr+u\otimes
g(^k\pi )+\,^ks\otimes g(u)+(^kS)g\text{ ,} 
\]

\noindent where 
\begin{equation}  \label{X.3.-1}
^k\pi =\,^G\overline{\pi }\text{ ,\thinspace \thinspace \thinspace
\thinspace \thinspace \thinspace \thinspace }^ks=\,^G\overline{s}\text{
,\thinspace \thinspace \thinspace \thinspace \thinspace \thinspace
\thinspace \thinspace }^kS=\,^G\overline{S}\text{ .}
\end{equation}

By the use of the relations 
\begin{equation}  \label{X.3.-2}
\begin{array}{c}
u(\rho _G+\frac 1e\cdot L\cdot k)=u\rho _G+L\cdot u(\frac 1e\cdot k)+\frac
1e\cdot k\cdot (uL)= \\ 
\\ 
=[\rho _{G/\alpha }+L\cdot (\frac 1e\cdot k)_{/\alpha }+\frac 1e\cdot k\cdot
L_{/\alpha }]\cdot u^\alpha = \\ 
\\ 
=[\rho _{G,j}+L\cdot (\frac 1e\cdot k)_{,j}+\frac 1e\cdot k\cdot
L_{,j}]\cdot u^j\text{ ,}
\end{array}
\end{equation}
\begin{equation}  \label{X.3.-3}
KrL=L_{/\alpha }\cdot e^\alpha =L_{,i}\cdot dx^i=\overline{\nabla }_{Kr}L%
\text{ ,}
\end{equation}
\begin{equation}  \label{X.3.-4}
\delta (L\cdot Kr)=KrL+L\cdot \delta Kr\text{ ,}
\end{equation}
\begin{equation}  \label{X.3.-5}
\begin{array}{c}
\delta (L\cdot Kr)=\frac 12\cdot [\nabla _{\overline{g}}(L.Kr)]g=(L.g_\alpha
^\beta )_{/\beta }\cdot e^\alpha =(L\cdot g_i^j)_{;j}\cdot dx^i= \\ 
=(L_{/\beta }\cdot g_\alpha ^\beta +L\cdot g_\alpha ^\beta \,_{/\beta
})\cdot e^\alpha =(L_{,i}+L\cdot g_i^j\,_{;j})\cdot dx^i\text{ ,}
\end{array}
\end{equation}
\begin{equation}  \label{X.3.-6}
\delta (u\otimes g(^G\overline{\pi }))=\delta u\cdot g(^G\overline{\pi }%
)+g(\nabla _u\,^G\overline{\pi })+(\nabla _ug)(^G\overline{\pi })\text{ ,}
\end{equation}
\begin{equation}  \label{X.3.-7}
\delta (^G\overline{s}\otimes g(u))=\delta ^G\overline{s}\cdot g(u)+g(\nabla
_{^G\overline{s}}u)+(\nabla _{^G\overline{s}}g)(u)\text{ ,}
\end{equation}
\[
\delta ((^G\overline{S})g)=(g_{\alpha \overline{\gamma }}\cdot \,^G\overline{%
S}\,^{\beta \gamma })_{/\beta }\cdot e^\alpha \text{ [see (\ref{X.2.-5})],} 
\]

\noindent $\delta G$ and $\overline{g}(\delta G)$ can be found in the forms 
\begin{equation}  \label{X.3.-8}
\begin{array}{c}
\delta G=(\rho _G+\frac 1e\cdot L\cdot k)\cdot g(a)+ \\ 
+[u(\rho _G+\frac 1e\cdot L\cdot k)+\,\,\,(\rho _G+\frac 1e\cdot L\cdot
k)\cdot \delta u+\delta ^G \overline{s}]\cdot g(u)- \\ 
-\,\,KrL-L\cdot \delta Kr+\delta u\cdot g(^G \overline{\pi })+g(\nabla _u\,^G%
\overline{\pi })+g(\nabla _{^G\overline{s}}u)+ \\ 
+\,\,(\rho _G+\frac 1e\cdot L\cdot k)\cdot (\nabla _ug)(u)+(\nabla _ug)(^G 
\overline{\pi })+(\nabla _{^G\overline{s}}g)(u)+ \\ 
+\,\,\delta ((^G\overline{S})g)\text{ ,}
\end{array}
\end{equation}
\begin{equation}  \label{X.3.-9}
\begin{array}{c}
\overline{g}(\delta G)=(\rho _G+\frac 1e\cdot L\cdot k)\cdot a+ \\ 
+[u(\rho _G+\frac 1e\cdot L\cdot k)+\,\,\,\,\,\,(\rho _G+\frac 1e\cdot
L\cdot k)\cdot \delta u+\delta ^G \overline{s}]\cdot u- \\ 
-\, \overline{g}(\,KrL)-L.\overline{g}(\delta Kr)+\delta u\cdot \,^G%
\overline{\pi }+\nabla _u\,^G\overline{\pi }+\nabla _{^G\overline{s}}u+ \\ 
+\,\,(\rho _G+\frac 1e\cdot L\cdot k)\cdot \overline{g}(\nabla _ug)(u)+%
\overline{g}(\nabla _ug)(^G\overline{\pi })+\overline{g}(\nabla _{^G%
\overline{s}}g)(u)+ \\ 
+\,\,\overline{g}(\delta ((^G\overline{S})g))\text{ .}
\end{array}
\end{equation}

In a co-ordinate basis $\delta G$ and $\overline{g}(\delta G)$ will have the
forms 
\begin{equation}  \label{X.3.-12}
\begin{array}{c}
G_i\,^j\,_{;j}=(\rho _G+\frac 1e\cdot L\cdot k)\cdot a_i+ \\ 
+\,\,[(\rho _G+\frac 1e\cdot L\cdot k)_{,j}\cdot u^j+(\rho _G+\frac 1e\cdot
L\cdot k)\cdot u^j\,_{;j}+\,^G \overline{s}^j\,_{;j}]\cdot u_i- \\ 
-L_{,i}-L\cdot g_{i\,;j}^j+u^j\,_{;j}\cdot \,^G \overline{\pi }_i+g_{i%
\overline{j}}\cdot (^G\overline{\pi }^j\,_{;k}\cdot u^k+u^j\,_{;k}\cdot \,^G%
\overline{s}^k)+ \\ 
+\,\,\,g_{ij;k}\cdot [(\rho _G+\frac 1e\cdot L\cdot k)\cdot u^{\overline{j}%
}\cdot u^k+\,^G\overline{\pi }^{\overline{j}}\cdot u^k+u^{\overline{j}}\cdot
\,^G\overline{s}^k]+ \\ 
+\,(g_{i\overline{k}}\cdot \,^G\overline{S}\,^{jk})_{;j}\text{ ,}
\end{array}
\end{equation}
\begin{equation}  \label{X.3.-13}
\begin{array}{c}
\overline{g}^{i\overline{k}}\cdot G_k\,^j\,_{;j}=(\rho _G+\frac 1e\cdot
L\cdot k)\cdot a^i+ \\ 
+\,\,[(\rho _G+\frac 1e\cdot L\cdot k)_{,j}\cdot u^j+(\rho _G+\frac 1e\cdot
L\cdot k)\cdot u^j\,_{;j}+\,^G \overline{s}^j\,_{;j}]\cdot u^i- \\ 
-L_{,j}\cdot g^{i \overline{j}}-L\cdot g^{i\overline{k}}\cdot
g_{k\,;j}^j+u^j\,_{;j}\cdot \,^G\overline{\pi }^i+\,^G\overline{\pi }%
^i\,_{;j}\cdot u^j+u^i\,_{;j}\cdot \,^G\overline{s}^j+ \\ 
+\,\,\,g^{i \overline{l}}\cdot g_{lj;k}\cdot [(\rho _G+\frac 1e\cdot L\cdot
k)\cdot u^{\overline{j}}\cdot u^k+\,^G\overline{\pi }^{\overline{j}}\cdot
u^k+u^{\overline{j}}\cdot \,^G\overline{s}^k]+ \\ 
+\,g^{i\overline{l}}\cdot (g_{l\overline{k}}\cdot \,^G\overline{S}%
\,^{jk})_{;j}\text{ .}
\end{array}
\end{equation}

On the grounds of the representations of $\delta G$ and $\overline{g}(\delta
G)$ the representation of the different energy-momentum tensors $\theta $, $%
_sT$ and $Q$ can be found.

The covariant divergency $\delta \theta $ of the generalized canonical
energy-momentum tensor (GC-EMT) $\theta $ can be written in the form 
\begin{equation}  \label{X.3.-14}
\begin{array}{c}
\delta \theta =(\rho _\theta +\frac 1e\cdot L\cdot k)\cdot g(a)+ \\ 
+[u(\rho _\theta +\frac 1e\cdot L\cdot k)+(\rho _\theta +\frac 1e\cdot
L\cdot k)\cdot \delta u+\delta ^\theta \overline{s}]\cdot g(u)- \\ 
-\,\,KrL-L.\delta Kr+\delta u\cdot g(^\theta \overline{\pi })+g(\nabla
_u\,^\theta \overline{\pi })+g(\nabla _{^\theta \overline{s}}u)+ \\ 
+\,\,(\rho _\theta +\frac 1e\cdot L\cdot k)\cdot (\nabla _ug)(u)+(\nabla
_ug)(^\theta \overline{\pi })+(\nabla _{^\theta \overline{s}}g)(u)+ \\ 
+\,\,\delta ((^\theta \overline{S})g)\text{ ,}
\end{array}
\end{equation}

\noindent or in the form 
\begin{equation}  \label{X.3.-15}
\begin{array}{c}
\overline{g}(\delta \theta )=(\rho _\theta +\frac 1e\cdot L\cdot k)\cdot a+
\\ 
+[u(\rho _\theta +\frac 1e\cdot L\cdot k)+(\rho _\theta +\frac 1e\cdot
L\cdot k)\cdot \delta u+\delta ^\theta \overline{s}]\cdot u- \\ 
-\, \overline{g}(\,KrL)-L\cdot \overline{g}(\delta Kr)+\delta u\cdot
\,^\theta \overline{\pi }+\nabla _u\,^\theta \overline{\pi }+\nabla
_{^\theta \overline{s}}u+ \\ 
+\,\,(\rho _\theta +\frac 1e\cdot L\cdot k). \overline{g}(\nabla _ug)(u)+%
\overline{g}(\nabla _ug)(^\theta \overline{\pi })+\overline{g}(\nabla
_{^\theta \overline{s}}g)(u)+ \\ 
+\,\,\overline{g}(\delta ((^\theta \overline{S})g))\text{ .}
\end{array}
\end{equation}

In a co-ordinate basis $\delta \theta $ and $\overline{g}(\delta \theta )$
will have the forms 
\begin{equation}  \label{X.3.-18}
\begin{array}{c}
\overline{\theta }_i\,^j\,_{;j}=(\rho _\theta +\frac 1e\cdot L\cdot k)\cdot
a_i+ \\ 
+\,\,[(\rho _\theta +\frac 1e\cdot L\cdot k)_{,j}\cdot u^j+\,\,\,(\rho
_\theta +\frac 1e\cdot L\cdot k)\cdot u^j\,_{;j}+\,^\theta \overline{s}%
^j\,_{;j}]\cdot u_i- \\ 
-L_{,i}-L\cdot g_{i\,;j}^j+u^j\,_{;j}\cdot \,^\theta \overline{\pi }_i+g_{i%
\overline{j}}\cdot (^\theta \overline{\pi }^j\,_{;k}\cdot
u^k+u^j\,_{;k}\cdot \,^\theta \overline{s}^k)+ \\ 
+\,\,\,g_{ij;k}\cdot [(\rho _\theta +\frac 1e\cdot L\cdot k)\cdot u^{%
\overline{j}}\cdot u^k+\,^\theta \overline{\pi }^{\overline{j}}\cdot u^k+u^{%
\overline{j}}\cdot \,^\theta \overline{s}^k]+ \\ 
+\,(g_{i\overline{k}}\cdot \,^\theta \overline{S}\,\,^{jk})_{;j}\text{ ,}
\end{array}
\end{equation}
\begin{equation}  \label{X.3.-19}
\begin{array}{c}
\overline{g}^{i\overline{k}}\cdot \overline{\theta }_k\,^j\,_{;j}=(\rho
_\theta +\frac 1e\cdot L\cdot k)\cdot a^i+ \\ 
+\,\,[(\rho _\theta +\frac 1e\cdot L\cdot k)_{,j}\cdot u^j+(\rho _\theta
+\frac 1e\cdot L\cdot k)\cdot u^j\,_{;j}+\,^\theta \overline{s}%
^j\,_{;j}]\cdot u^i- \\ 
-L_{,j}\cdot g^{i \overline{j}}-L\cdot g^{i\overline{k}}\cdot
g_{k\,;j}^j+u^j\,_{;j}\cdot \,^\theta \overline{\pi }^i+\,^\theta \overline{%
\pi }^i\,_{;j}\cdot u^j+u^i\,_{;j}\cdot \,^\theta \overline{s}^j+ \\ 
+\,\,\,g^{i \overline{l}}\cdot g_{lj;k}\cdot [(\rho _\theta +\frac 1e\cdot
L\cdot k)\cdot u^{\overline{j}}\cdot u^k+\,^\theta \overline{\pi }^{%
\overline{j}}\cdot u^k+u^{\overline{j}}\cdot \,^\theta \overline{s}^k]+ \\ 
+\,g^{i\overline{l}}\cdot (g_{l\overline{k}}\cdot \,^\theta \overline{S}%
\,^{jk})_{;j}\text{ .}
\end{array}
\end{equation}

The covariant divergency $\delta _sT$ of the symmetric energy-momentum
tensor of Belinfante (S-EMT-B) $_sT$, represented in the form 
\[
_sT=(\rho _T+\frac 1e\cdot L\cdot k).u\otimes g(u)-L\cdot Kr+u\otimes g(^T%
\overline{\pi })+\,^T\overline{s}\otimes g(u)+(^T\overline{S})g\text{ ,} 
\]

\noindent can be found in the form 
\begin{equation}  \label{X.3.-20}
\begin{array}{c}
\delta _sT=(\rho _T+\frac 1e\cdot L\cdot k).g(a)+ \\ 
+[u(\rho _T+\frac 1e\cdot L\cdot k)+(\rho _T+\frac 1e\cdot L\cdot k)\cdot
\delta u+\delta ^T \overline{s}]\cdot g(u)- \\ 
-\,\,KrL-L\cdot \delta Kr+\delta u\cdot g(^T \overline{\pi })+g(\nabla _u\,^T%
\overline{\pi })+g(\nabla _{^T\overline{s}}u)+ \\ 
+\,\,(\rho _T+\frac 1e\cdot L\cdot k)\cdot (\nabla _ug)(u)+(\nabla _ug)(^T 
\overline{\pi })+(\nabla _{^T\overline{s}}g)(u)+ \\ 
+\,\,\delta ((^T\overline{S})g)\text{ ,}
\end{array}
\end{equation}

\noindent or in the form 
\begin{equation}  \label{X.3.-21}
\begin{array}{c}
\overline{g}(\delta _sT)=(\rho _T+\frac 1e\cdot L\cdot k)\cdot a+ \\ 
+[u(\rho _T+\frac 1e\cdot L\cdot k)+\,\,\,\,\,\,(\rho _T+\frac 1e\cdot
L\cdot k)\cdot \delta u+\delta ^T \overline{s}]\cdot u- \\ 
-\, \overline{g}(\,KrL)-L\cdot \overline{g}(\delta Kr)+\delta u\cdot \,^T%
\overline{\pi }+\nabla _u\,^T\overline{\pi }+\nabla _{^T\overline{s}}u+ \\ 
+\,\,(\rho _T+\frac 1e\cdot L\cdot k)\cdot \overline{g}(\nabla _ug)(u)+%
\overline{g}(\nabla _ug)(^T\overline{\pi })+\overline{g}(\nabla _{^T%
\overline{s}}g)(u)+ \\ 
+\,\,\overline{g}(\delta ((^T\overline{S})g))\text{ .}
\end{array}
\end{equation}

In a co-ordinate basis $\delta _sT$ and $\overline{g}(\delta _sT)$ will have
the forms 
\begin{equation}  \label{X.3.-24}
\begin{array}{c}
_sT_i\,^j\,_{;j}=(\rho _T+\frac 1e\cdot L\cdot k)\cdot a_i+ \\ 
+\,\,[(\rho _T+\frac 1e\cdot L\cdot k)_{,j}\cdot u^j+\,\,\,\,\,\,(\rho
_T+\frac 1e\cdot L\cdot k)\cdot u^j\,_{;j}+\,^T \overline{s}^j\,_{;j}]\cdot
u_i- \\ 
-L_{,i}-L\cdot g_{i\,;j}^j+u^j\,_{;j}\cdot \,^T \overline{\pi }_i+g_{i%
\overline{j}}\cdot (^T\overline{\pi }^j\,_{;k}\cdot u^k+u^j\,_{;k}\cdot \,^T%
\overline{s}^k)+ \\ 
+\,\,\,g_{ij;k}\cdot [(\rho _T+\frac 1e\cdot L\cdot k)\cdot u^{\overline{j}%
}\cdot u^k+\,^T\overline{\pi }^{\overline{j}}\cdot u^k+u^{\overline{j}}\cdot
\,^T\overline{s}^k]+ \\ 
+\,(g_{i\overline{k}}\cdot \,^T\overline{S}\,^{jk})_{;j}\text{ ,}
\end{array}
\end{equation}
\begin{equation}  \label{X.3.-25}
\begin{array}{c}
\overline{g}^{i\overline{k}}\cdot \,_sT_k\,^j\,_{;j}=(\rho _T+\frac 1e\cdot
L\cdot k)\cdot a^i+ \\ 
+\,\,[(\rho _T+\frac 1e\cdot L\cdot k)_{;j}\cdot u^j+(\rho _T+\frac 1e\cdot
L\cdot k)\cdot u^j\,_{;j}+\,^T \overline{s}^j\,_{;j}]\cdot u^i- \\ 
-L_{,j}\cdot g^{i \overline{j}}-L\cdot g^{i\overline{k}}\cdot
g_{k\,;j}^j+u^j\,_{;j}\cdot \,^T\overline{\pi }^i+\,^T\overline{\pi }%
^i\,_{;j}\cdot u^j+u^i\,_{;j}\cdot \,^T\overline{s}^j+ \\ 
+\,\,\,g^{i \overline{l}}\cdot g_{lj;k}\cdot [(\rho _T+\frac 1e\cdot L\cdot
k)\cdot u^{\overline{j}}\cdot u^k+\,^T\overline{\pi }^{\overline{j}}\cdot
u^k+u^{\overline{j}}\cdot \,^T\overline{s}^k]+ \\ 
+\,g^{i\overline{l}}\cdot (g_{l\overline{k}}\cdot \,^T\overline{S}%
\,^{jk})_{;j}\text{ .}
\end{array}
\end{equation}

The covariant divergency $\delta Q$ of the variational energy-momentum
tensor of Euler-Lagrange (V-EMT-EL) $Q$, represented in the form 
\[
Q=-\,\rho _Q\cdot u\otimes g(u)-u\otimes g(^Q\pi )-\,^Qs\otimes g(u)-(^QS)g%
\text{ ,} 
\]

\noindent follows in the form 
\begin{equation}  \label{X.3.-26}
\begin{array}{c}
\delta Q=-\,\rho _Q\cdot g(a)-(u\rho _Q+\rho _Q\cdot \delta u+\delta
^Qs)\cdot g(u)- \\ 
-\,\,\delta u\cdot g(^Q\pi )-g(\nabla _u\,^Q\pi )-g(\nabla _{^Qs}u)- \\ 
-\,\,\rho _Q\cdot (\nabla _ug)(u)-(\nabla _ug)(^Q\pi )-(\nabla
_{^Qs}g)(u)-\,\delta ((^QS)g)\text{ ,}
\end{array}
\end{equation}

\noindent or in the form 
\begin{equation}  \label{X.3.-27}
\begin{array}{c}
\overline{g}(\delta Q)=-\,\rho _Q\cdot a-(u\rho _Q+\rho _Q\cdot \delta
u+\delta ^Qs)\cdot u- \\ 
-\,\,\delta u\cdot \,^Q\pi -\nabla _u\,^Q\pi -\nabla _{^Qs}u- \\ 
-\,\,\rho _Q\cdot \overline{g}(\nabla _ug)(u)-\overline{g}(\nabla _ug)(^Q\pi
)-\overline{g}(\nabla _{^Qs}g)(u)-\,\overline{g}(\delta ((^QS)g))\text{ .}
\end{array}
\end{equation}

In a co-ordinate basis $\delta Q$ and $\overline{g}(\delta Q)$ will have the
forms 
\begin{equation}  \label{X.3.-30}
\begin{array}{c}
\overline{Q}_i\,^j\,_{;j}=-\,\rho _Q\cdot a_i-(\rho _Q{}_{,j}\cdot u^j+\rho
_Q\cdot u^j\,_{;j}+\,^Qs^j\,_{;j})\cdot u_i- \\ 
-\,u^j\,_{;j}\cdot \,^Q\pi _i-g_{i \overline{j}}\cdot (^Q\pi ^j\,_{;k}\cdot
u^k+u^j\,_{;k}\cdot \,^Qs^k)- \\ 
-\,\,\,g_{ij;k}\cdot (\rho _Q\cdot u^{\overline{j}}\cdot u^k+\,^Q\pi ^{%
\overline{j}}\cdot u^k+u^{\overline{j}}\cdot \,^Qs^k)-(g_{i\overline{k}%
}\cdot \,^QS^{jk})_{;j}\text{ ,}
\end{array}
\end{equation}
\begin{equation}  \label{X.3.-31}
\begin{array}{c}
\overline{g}^{i\overline{k}}\cdot \overline{Q}_k\,^j\,_{;j}=-\,\rho _Q\cdot
a^i-(\rho _{Q,j}\cdot u^j+\rho _Q\cdot u^j\,_{;j}+\,^Qs^j\,_{;j})\cdot u^i-
\\ 
-\,\,u^j\,_{;j}\cdot \,^Q\pi ^i-\,^Q\pi ^i\,_{;j}\cdot u^j-u^i\,_{;j}\cdot
\,^Qs^j- \\ 
-\,\,\,g^{i\overline{l}}\cdot g_{lj;k}\cdot (\rho _Q\cdot u^{\overline{j}%
}\cdot u^k+\,^Q\pi ^{\overline{j}}\cdot u^k+u^{\overline{j}}\cdot
\,^Qs^k]-g^{i\overline{l}}\cdot (g_{l\overline{k}}\cdot \,^QS^{jk})_{;j}%
\text{ .}
\end{array}
\end{equation}

\section{Covariant divergency of the energy-momentum tensors and the
momentum density}

In the previous chapter the energy-momentum tensors are represented by the
use of the projective metrics $h^u$, $h_u$ and the momentum density $p$. For
this type of representation of a tensor field $G$ of the type 1., which has
the form 
\[
G=u\otimes g(p_G)+\,^Gs\otimes g(u)+(^GS)g\text{ ,} 
\]

\noindent the covariant divergency $\delta G$ can be calculated in the form 
\begin{equation}  \label{X.4.-1}
\begin{array}{c}
\delta G=g(\nabla _up_G)+\delta u\cdot g(p_G)+\delta ^Gs\cdot g(u)+g(\nabla
_{^Gs}u)+(\nabla _ug)(p_G)+ \\ 
+\,\,\,(\nabla _{^Gs}g)(u)+\delta ((^GS)g)\text{ ,}
\end{array}
\end{equation}

\noindent or in the form 
\begin{equation}  \label{X.4.-2}
\begin{array}{c}
\overline{g}(\delta G)=\nabla _up_G+\delta u\cdot p_G+\delta ^Gs\cdot
u+\nabla _{^Gs}u+\overline{g}(\nabla _ug)(p_G)+ \\ 
+\,\,\,\overline{g}(\nabla _{^Gs}g)(u)+\overline{g}(\delta ((^GS)g))\text{ .}
\end{array}
\end{equation}

In a co-ordinate basis $\delta G$ and $\overline{g}(\delta G)$ will have the
forms 
\begin{equation}  \label{X.4.-5}
\begin{array}{c}
G_i\,^j\,_{;j}=g_{i \overline{j}}\cdot (p_{G;k}^j\cdot u^k+u^k\,_{;k}\cdot
p_G^j+\,^Gs^k\,_{;k}\cdot u^j+u^j\,_{;k}\cdot \,^Gs^k)+ \\ 
+\,g_{ij;k}\cdot (p_G^{\overline{j}}\cdot u^k+u^{\overline{j}}\cdot
\,^Gs^k)+(g_{i\overline{k}}\cdot \,^GS^{jk})_{;j}\text{ ,}
\end{array}
\end{equation}
\begin{equation}  \label{X.4.-6}
\begin{array}{c}
g^{i \overline{k}}\cdot G_k\,^j\,_{;j}=p_{G;j}^i\cdot u^j+u^j\,_{;j}\cdot
p_G^i+\,^Gs^j\,_{;j}\cdot u^i+u^i\,_{;j}\cdot \,^Gs^j+ \\ 
+\,\overline{g}^{i\overline{l}}\cdot g_{lj;k}\cdot (p_G^{\overline{j}}\cdot
u^k+u^{\overline{j}}\cdot \,^Gs^k)+g^{i\overline{l}}\cdot (g_{l\overline{k}%
}\cdot \,^GS^{jk})_{;j}\text{ .}
\end{array}
\end{equation}

On the other side, the covariant divergency of $(G)\overline{g}$ can be
found. $(G)\overline{g}$ is represented in the form 
\begin{equation}  \label{X.4.-7}
(G)\overline{g}=u\otimes p_G+\,^Gs\otimes u+\,^GS=G^{\alpha \beta }\cdot
e_\alpha \otimes e_\beta =G^{ij}\cdot \partial _i\otimes \partial _j\text{ .}
\end{equation}

\noindent $\delta ((G)\overline{g})$ will have the form 
\begin{equation}  \label{X.4.-8}
\delta ((G)\overline{g})=\nabla _u\,^Gs+\nabla _{p_G}u+\delta p_G\cdot
u+\delta u\cdot \,^Gs+\delta ^GS\text{ ,}
\end{equation}
\begin{equation}  \label{X.4.-9}
G^{\alpha \beta }\,_{/\beta }=\,^Gs^\alpha \,_{/\beta }\cdot u^\beta
+u^\alpha \,_{/\beta }\cdot p_G^\beta +p_G^\beta \,_{/\beta }\cdot u^\alpha
+u^\beta \,_{/\beta }\cdot \,^Gs^\alpha +\,^GS^{\alpha \beta }\,_{/\beta }%
\text{ ,}
\end{equation}
\begin{equation}  \label{X.4.-10}
G^{ij}\,_{;j}=\,^Gs^i\,_{;j}\cdot u^j+u^i\,_{;j}\cdot
p_G^j+p_G^j\,_{;j}\cdot u^i+u^j\,_{;j}\cdot \,^Gs^i+\,^GS^{ij}\,_{;j}\text{ ,%
}
\end{equation}
\[
G^{\alpha \beta }=G_\gamma \,^\alpha \cdot g^{\overline{\gamma }\beta }\text{
,\thinspace \thinspace \thinspace \thinspace \thinspace \thinspace
\thinspace \thinspace \thinspace }G^{ij}=G_k\,^i\cdot g^{\overline{k}j}\text{
.} 
\]

The covariant divergency $\delta \theta $ of the generalized canonical
energy-momentum tensor $\theta $ follows from (\ref{X.4.-1}) and (\ref
{X.4.-2}) in the form 
\begin{equation}  \label{X.4.-11}
\begin{array}{c}
\delta \theta =g(\nabla _up_\theta )+\delta u\cdot g(p_\theta )+\delta
^\theta s\cdot g(u)+g(\nabla _{^\theta s}u)+(\nabla _ug)(p_\theta )+ \\ 
+\,\,\,(\nabla _{^\theta s}g)(u)+\delta ((^\theta S)g)\text{ ,}
\end{array}
\end{equation}

\noindent or in the form 
\begin{equation}  \label{X.4.-12}
\begin{array}{c}
\overline{g}(\delta \theta )=\nabla _up_\theta +\delta u\cdot p_\theta
+\delta ^\theta s\cdot u+\nabla _{^\theta s}u+\overline{g}(\nabla
_ug)(p_\theta )+ \\ 
+\,\,\,\overline{g}(\nabla _{^\theta s}g)(u)+\overline{g}(\delta ((^\theta
S)g))\text{ .}
\end{array}
\end{equation}

In a co-ordinate basis $\delta \theta $ and $\overline{g}(\delta \theta )$
will have the forms 
\begin{equation}  \label{X.4.-15}
\begin{array}{c}
\overline{\theta }_i\,^j\,_{;j}=g_{i\overline{j}}\cdot (p_{\theta ;k}^j\cdot
u^k+u^k\,_{;k}\cdot p_\theta ^j+\,^\theta s^k\,_{;k}\cdot
u^j+u^j\,_{;k}\cdot \,^\theta s^k)+ \\ 
+\,g_{ij;k}\cdot (p_\theta ^{\overline{j}}\cdot u^k+u^{\overline{j}}\cdot
\,^\theta s^k)+(g_{i\overline{k}}\cdot \,^\theta S^{jk})_{;j}\text{ ,}
\end{array}
\end{equation}
\begin{equation}  \label{X.4.-16}
\begin{array}{c}
g^{i \overline{k}}\cdot \overline{\theta }_k\,^j\,_{;j}=p_{\theta ;j}^i\cdot
u^j+u^j\,_{;j}\cdot p_\theta ^i+\,^\theta s^j\,_{;j}\cdot
u^i+u^i\,_{;j}\cdot \,^\theta s^j+ \\ 
+\,\overline{g}^{i\overline{l}}\cdot g_{lj;k}\cdot (p_\theta ^{\overline{j}%
}\cdot u^k+u^{\overline{j}}\cdot \,^\theta s^k)+g^{i\overline{l}}\cdot (g_{l%
\overline{k}}\cdot \,^\theta S^{jk})_{;j}\text{ .}
\end{array}
\end{equation}

The covariant divergency $\delta _sT$ of the symmetric energy-momentum
tensor of Belinfante $_sT$ can be written in the form 
\begin{equation}  \label{X.4.-17}
\begin{array}{c}
\delta _sT=g(\nabla _up_T)+\delta u\cdot g(p_T)+\delta ^Ts\cdot
g(u)+g(\nabla _{^Ts}u)+(\nabla _ug)(p_T)+ \\ 
+\,\,\,(\nabla _{^Ts}g)(u)+\delta ((^TS)g)\text{ ,}
\end{array}
\end{equation}

\noindent or in the form 
\begin{equation}  \label{X.4.-18}
\begin{array}{c}
\overline{g}(\delta _sT)=\nabla _up_T+\delta u\cdot p_T+\delta ^Ts\cdot
u+\nabla _{^Ts}u+\overline{g}(\nabla _ug)(p_T)+ \\ 
+\,\,\,\overline{g}(\nabla _{^Ts}g)(u)+\overline{g}(\delta ((^TS)g))\text{ .}
\end{array}
\end{equation}

In a co-ordinate basis $\delta _sT$ and $\overline{g}(\delta _sT)$ will have
the forms 
\begin{equation}  \label{X.4.-21}
\begin{array}{c}
_sT_i\,^j\,_{;j}=g_{i \overline{j}}\cdot (p_{T;k}^j\cdot u^k+u^k\,_{;k}\cdot
p_T^j+\,^Ts^k\,_{;k}\cdot u^j+u^j\,_{;k}\cdot \,^Ts^k)+ \\ 
+\,g_{ij;k}\cdot (p_T^{\overline{j}}\cdot u^k+u^{\overline{j}}\cdot
\,^Ts^k)+(g_{i\overline{k}}\cdot \,^TS^{jk})_{;j}\text{ ,}
\end{array}
\end{equation}
\begin{equation}  \label{X.4.-22}
\begin{array}{c}
g^{i \overline{k}}\cdot \,_sT_k\,^j\,_{;j}=p_{T;j}^i\cdot
u^j+u^j\,_{;j}\cdot p_T^i+\,^Ts^j\,_{;j}\cdot u^i+u^i\,_{;j}\cdot \,^Ts^j+
\\ 
+\,\overline{g}^{i\overline{l}}\cdot g_{lj;k}\cdot (p_T^{\overline{j}}\cdot
u^k+u^{\overline{j}}\cdot \,^Ts^k)+g^{i\overline{l}}\cdot (g_{l\overline{k}%
}\cdot \,^TS^{jk})_{;j}\text{ .}
\end{array}
\end{equation}

The covariant divergency $\delta Q$ of the variational energy-momentum
tensor of Euler-Lagrange $Q$ can be found in the forms 
\begin{equation}  \label{X.4.-23}
\begin{array}{c}
\delta Q=-\,\,g(\nabla _up_Q)-\delta u\cdot g(p_Q)-\delta ^Qs\cdot
g(u)-g(\nabla _{^Qs}u)-(\nabla _ug)(p_Q)- \\ 
-\,\,\,(\nabla _{^Qs}g)(u)-\delta ((^QS)g)\text{ ,}
\end{array}
\end{equation}
\begin{equation}  \label{X.4.-24}
\begin{array}{c}
\overline{g}(\delta Q)=-\,\,\nabla _up_Q-\delta u\cdot p_Q-\delta ^Qs\cdot
u-\nabla _{^Qs}u-\overline{g}(\nabla _ug)(p_Q)- \\ 
-\,\,\,\overline{g}(\nabla _{^Qs}g)(u)-\overline{g}(\delta ((^QS)g))\text{ .}
\end{array}
\end{equation}

In a co-ordinate basis $\delta Q$ and $\overline{g}(\delta Q)$ will have the
forms 
\begin{equation}  \label{X.4.-27}
\begin{array}{c}
\overline{Q}_i\,^j\,_{;j}=-\,\,\,g_{i\overline{j}}\cdot (p_{Q;k}^j\cdot
u^k+u^k\,_{;k}\cdot p_Q^j+\,^Qs^k\,_{;k}\cdot u^j+u^j\,_{;k}\cdot \,^Qs^k)-
\\ 
-\;\,g_{ij;k}\cdot (p_Q^{\overline{j}}\cdot u^k+u^{\overline{j}}\cdot
\,^Qs^k)+(g_{i\overline{k}}\cdot \,^QS^{jk})_{;j}\text{ ,}
\end{array}
\end{equation}
\begin{equation}  \label{X.4.-28}
\begin{array}{c}
g^{i \overline{k}}\cdot \overline{Q}_k\,^j\,_{;j}=-\,\,\,p_{Q;j}^i\cdot
u^j-u^j\,_{;j}\cdot p_Q^i-\,^Qs^j\,_{;j}\cdot u^i-u^i\,_{;j}\cdot \,^Qs^j-
\\ 
-\,\,\,\overline{g}^{i\overline{l}}\cdot g_{lj;k}\cdot (p_Q^{\overline{j}%
}\cdot u^k+u^{\overline{j}}\cdot \,^Qs^k)-g^{i\overline{l}}\cdot (g_{l%
\overline{k}}\cdot \,^QS^{jk})_{;j}\text{ .}
\end{array}
\end{equation}

\section{Covariant divergency of the energy-momentum tensors and the energy
flux density}

In the previous chapter the notion of energy flux density $e_G$ has been
introduced. By means of it a mixed tensor field $G$ of second rank and of
type 1. can be represented in the form 
\[
G=\frac 1e\cdot e_G\otimes g(u)+u\otimes g(^G\pi )+(^GS)g\text{ .} 
\]

The covariant divergency $\delta G$ of a given through $e_G$ tensor field $G$
can be found by the use of the relation 
\begin{equation}  \label{X.5.-1}
\delta (\frac 1e\cdot e_G\otimes g(u))=[e_G(\frac 1e)+\frac 1e\cdot \delta
e_G]\cdot g(u)+\frac 1e\cdot [(\nabla _{e_G}g)(u)+g(\nabla _{e_G}u)]
\end{equation}

\noindent and the property (\ref{X.1.-10}) of $\delta $ in the form 
\begin{equation}  \label{X.5.-2}
\begin{array}{c}
\delta G=g(\nabla _u\,^G\pi )+\frac 1e\cdot g(\nabla _{e_G}u)+[e_G(\frac
1e)+\frac 1e\cdot \delta e_G]\cdot g(u)+\delta u\cdot g(^G\pi )+ \\ 
+\,\,\frac 1e\cdot (\nabla _{e_G}g)(u)+(\nabla _ug)(^G\pi )+\delta ((^GS)g)%
\text{ ,}
\end{array}
\end{equation}

\noindent or in the form 
\begin{equation}  \label{X.5.-3}
\begin{array}{c}
\overline{g}(\delta G)=\nabla _u\,^G\pi +\frac 1e\cdot \nabla
_{e_G}u+[e_G(\frac 1e)+\frac 1e\cdot \delta e_G]\cdot u+\delta u\cdot
\,^G\pi + \\ 
+\,\,\frac 1e\cdot \overline{g}(\nabla _{e_G}g)(u)+\overline{g}(\nabla
_ug)(^G\pi )+\overline{g}(\delta ((^GS)g))\text{ .}
\end{array}
\end{equation}

On the other side, the covariant divergency of $(G)\overline{g}$ can be
found in the form 
\begin{equation}  \label{X.5.-6}
\delta ((G)\overline{g})=\frac 1e\cdot \nabla _ue_G+\nabla _{^G\pi
}u+[u(\frac 1e)+\frac 1e\cdot \delta u]\cdot e_G+\delta ^G\pi \cdot u+\delta
(^GS)\text{ .}
\end{equation}

In a co-ordinate basis $\delta G$ and $\overline{g}(\delta G)$ will have the
forms 
\begin{equation}  \label{X.5.-9}
\begin{array}{c}
G_i\,^j\,_{;j}=g_{i \overline{j}}\cdot \{^G\pi ^j\,_{;k}\cdot u^k+\frac
1e\cdot u^j\,_{;k}\cdot e_G^k+[(\frac 1e)_{,k}\cdot e_G^k+\frac 1e\cdot
e_{G;k}^k]\cdot u^j+u^k\,_{;k}\cdot \,^G\pi ^j\}+ \\ 
+\,g_{ij;k}\cdot [\frac 1e\cdot u^{\overline{j}}\cdot e_G^k+\,^G\pi ^{%
\overline{j}}\cdot u^k]+(g_{i\overline{k}}\cdot \,^GS^{jk})_{;j}\text{ ,}
\end{array}
\end{equation}
\begin{equation}  \label{X.5.-10}
\begin{array}{c}
g^{i \overline{k}}\cdot G_k\,^j\,_{;j}=\,^G\pi ^i\,_{;j}\cdot u^j+\frac
1e\cdot u^i\,_{;j}\cdot e_G^j+[(\frac 1e)_{,j}\cdot e_G^j+\frac 1e\cdot
e_{G;j}^j]\cdot u^i+u^j\,_{;j}\cdot \,^G\pi ^i+ \\ 
+\,g^{i\overline{l}}\cdot g_{lj;k}\cdot [\frac 1e\cdot u^{\overline{j}}\cdot
e_G^k+\,^G\pi ^{\overline{j}}\cdot u^k]+g^{i\overline{l}}\cdot (g_{l%
\overline{k}}\cdot \,^GS^{jk})_{;j}\text{ , \thinspace \thinspace \thinspace
\thinspace \thinspace \thinspace \thinspace }(\frac 1e)_{;j}=(\frac 1e)_{,j}%
\text{ .}
\end{array}
\end{equation}

The different possibilities for a representation of the covariant divergency
of the energy-momentum tensors are related to the different possibilities
for a representation of the first covariant Noether identity for a given
Lagrangian system with its corresponding energy-momentum tensors.

The covariant divergency $\delta \theta $ of the generalized canonical
energy-momentum tensor $\theta $ can be written by the use of the energy
flux density $e_\theta $ in the form 
\begin{equation}  \label{X.5.-11}
\begin{array}{c}
\delta \theta =g(\nabla _u\,^\theta \pi )+\frac 1e\cdot g(\nabla _{e_\theta
}u)+[e_\theta (\frac 1e)+\frac 1e\cdot \delta e_\theta ]\cdot g(u)+\delta
u\cdot g(^\theta \pi )+ \\ 
+\,\,\frac 1e\cdot (\nabla _{e_\theta }g)(u)+(\nabla _ug)(^\theta \pi
)+\delta ((^\theta S)g)\text{ ,}
\end{array}
\end{equation}

\noindent or in the form 
\begin{equation}  \label{X.5.-12}
\begin{array}{c}
\overline{g}(\delta \theta )=\nabla _u\,^\theta \pi +\frac 1e\cdot \nabla
_{e_\theta }u+[e_\theta (\frac 1e)+\frac 1e\cdot \delta e_\theta ]\cdot
u+\delta u\cdot \,^\theta \pi + \\ 
+\,\,\frac 1e\cdot \overline{g}(\nabla _{e_\theta }g)(u)+\overline{g}(\nabla
_ug)(^\theta \pi )+\overline{g}(\delta ((^\theta S)g))\text{ .}
\end{array}
\end{equation}

In a co-ordinate basis $\delta \theta $ and $\overline{g}(\delta \theta )$
will have the forms 
\begin{equation}  \label{X.5.-15}
\begin{array}{c}
\overline{\theta }_i\,^j\,_{;j}=g_{i\overline{j}}\cdot \{^\theta \pi
^j\,_{;k}\cdot u^k+\frac 1e\cdot u^j\,_{;k}\cdot e_\theta ^k+[(\frac
1e)_{,k}\cdot e_\theta ^k+\frac 1e\cdot e_{\theta ;k}^k]\cdot
u^j+u^k\,_{;k}\cdot \,^\theta \pi ^j\}+ \\ 
+\,g_{ij;k}\cdot [\frac 1e\cdot u^{\overline{j}}\cdot e_\theta ^k+\,^\theta
\pi ^{\overline{j}}\cdot u^k]+(g_{i\overline{k}}\cdot \,^\theta S^{jk})_{;j}%
\text{ ,}
\end{array}
\end{equation}
\begin{equation}  \label{X.5.-16}
\begin{array}{c}
g^{i \overline{k}}\cdot \overline{\theta }_k\,^j\,_{;j}=\,^\theta \pi
^i\,_{;j}\cdot u^j+\frac 1e\cdot u^i\,_{;j}\cdot e_\theta ^j+[(\frac
1e)_{,j}\cdot e_\theta ^j+\frac 1e\cdot e_{\theta ;j}^j]\cdot
u^i+u^j\,_{;j}\cdot \,^\theta \pi ^i+ \\ 
+\,g^{i\overline{l}}\cdot g_{lj;k}\cdot [\frac 1e\cdot u^{\overline{j}}\cdot
e_\theta ^k+\,^\theta \pi ^{\overline{j}}\cdot u^k]+g^{i\overline{l}}\cdot
(g_{l\overline{k}}\cdot \,^\theta S^{jk})_{;j}\text{ .}
\end{array}
\end{equation}

The covariant divergency $\delta _sT$ of the symmetric energy-momentum
tensor of Belinfante $_sT$ is found by the use of $e_T$ in the form 
\begin{equation}  \label{X.5.-17}
\begin{array}{c}
\delta _sT=g(\nabla _u\,^T\pi )+\frac 1e\cdot g(\nabla _{e_T}u)+[e_T(\frac
1e)+\frac 1e\cdot \delta e_T]\cdot g(u)+\delta u\cdot g(^T\pi )+ \\ 
+\,\,\frac 1e\cdot (\nabla _{e_T}g)(u)+(\nabla _ug)(^T\pi )+\delta ((^TS)g)%
\text{ ,}
\end{array}
\end{equation}

\noindent or in the form 
\begin{equation}  \label{X.5.-18}
\begin{array}{c}
\overline{g}(\delta _sT)=\nabla _u\,^T\pi +\frac 1e\cdot \nabla
_{e_T}u+[e_T(\frac 1e)+\frac 1e\cdot \delta e_T]\cdot u+\delta u\cdot ^T\pi +
\\ 
+\,\,\frac 1e\cdot \overline{g}(\nabla _{e_T}g)(u)+\overline{g}(\nabla
_ug)(^T\pi )+\overline{g}(\delta ((^TS)g))\text{ .}
\end{array}
\end{equation}

In a co-ordinate basis $\delta _sT$ and $\overline{g}(\delta _sT)$ will have
the forms 
\begin{equation}  \label{X.5.-21}
\begin{array}{c}
_sT_i\,^j\,_{;j}=g_{i \overline{j}}\cdot \{^T\pi ^j\,_{;k}\cdot u^k+\frac
1e\cdot u^j\,_{;k}\cdot e_T^k+[(\frac 1e)_{,k}\cdot e_T^k+\frac 1e\cdot
e_{T;k}^k]\cdot u^j+u^k\,_{;k}\cdot \,^T\pi ^j\}+ \\ 
+\,g_{ij;k}\cdot [\frac 1e\cdot u^{\overline{j}}\cdot e_T^k+\,^T\pi ^{%
\overline{j}}\cdot u^k]+(g_{i\overline{k}}\cdot \,^TS^{jk})_{;j}\text{ ,}
\end{array}
\end{equation}
\begin{equation}  \label{X.5.-22}
\begin{array}{c}
g^{i \overline{k}}\cdot \,_sT_k\,^j\,_{;j}=\,^T\pi ^i\,_{;j}\cdot u^j+\frac
1e\cdot u^i\,_{;j}\cdot e_T^j+[(\frac 1e)_{,j}\cdot e_T^j+\frac 1e\cdot
e_{T;j}^j]\cdot u^i+u^j\,_{;j}\cdot \,^T\pi ^i+ \\ 
+\,g^{i\overline{l}}\cdot g_{lj;k}\cdot [\frac 1e\cdot u^{\overline{j}}\cdot
e_T^k+\,^T\pi ^{\overline{j}}\cdot u^k]+g^{i\overline{l}}\cdot (g_{l%
\overline{k}}\cdot \,^TS^{jk})_{;j}\text{ .}
\end{array}
\end{equation}

The covariant divergency $\delta Q$ of the variational energy-momentum
tensor of Euler-Lagrange $Q$ can be found by means of $e_Q$ in the forms 
\begin{equation}  \label{X.5.-23}
\begin{array}{c}
\delta Q=-\,\,\,g(\nabla _u\,^Q\pi )-\frac 1e\cdot g(\nabla
_{e_Q}u)-[e_Q(\frac 1e)+\frac 1e\cdot \delta e_Q]\cdot g(u)-\delta u\cdot
g(^Q\pi )- \\ 
-\,\,\frac 1e\cdot (\nabla _{e_Q}g)(u)-(\nabla _ug)(^Q\pi )-\delta ((^QS)g)%
\text{ ,}
\end{array}
\end{equation}

\begin{equation}  \label{X.5.-24}
\begin{array}{c}
\overline{g}(\delta Q)=-\,\,\nabla _u\,^Q\pi -\frac 1e\cdot \nabla
_{e_Q}u-[e_Q(\frac 1e)+\frac 1e\cdot \delta e_Q].u-\delta u\cdot ^Q\pi - \\ 
-\,\,\frac 1e\cdot \overline{g}(\nabla _{e_Q}g)(u)-\overline{g}(\nabla
_ug)(^Q\pi )-\overline{g}(\delta ((^QS)g))\text{ .}
\end{array}
\end{equation}

In a co-ordinate basis $\delta Q$ and $\overline{g}(\delta Q)$ will have the
forms 
\begin{equation}  \label{X.5.-27}
\begin{array}{c}
\overline{Q}_i\,^j\,_{;j}=-\,\,\,g_{i\overline{j}}\cdot \{^Q\pi
^j\,_{;k}\cdot u^k+\frac 1e\cdot u^j\,_{;k}\cdot e_Q^k+[(\frac 1e)_{,k}\cdot
e_Q^k+\frac 1e\cdot e_{Q;k}^k]\cdot u^j+u^k\,_{;k}\cdot \,^Q\pi ^j\}- \\ 
-\,g_{ij;k}\cdot [\frac 1e\cdot u^{\overline{j}}\cdot e_Q^k+\,^Q\pi ^{%
\overline{j}}\cdot u^k]-(g_{i\overline{k}}\cdot \,^QS^{jk})_{;j}\text{ ,}
\end{array}
\end{equation}
\begin{equation}  \label{X.5.-28}
\begin{array}{c}
g^{i \overline{k}}\cdot \overline{Q}_k\,^j\,_{;j}=\,-\;^Q\pi ^i\,_{;j}\cdot
u^j-\frac 1e\cdot u^i\,_{;j}\cdot e_Q^j-[(\frac 1e)_{,j}\cdot e_Q^j+\frac
1e\cdot e_{Q;j}^j]\cdot u^i-u^j\,_{;j}\cdot \,^Q\pi ^i- \\ 
-\,\,g^{i\overline{l}}\cdot g_{lj;k}\cdot [\frac 1e\cdot u^{\overline{j}%
}\cdot e_Q^k+\,^Q\pi ^{\overline{j}}\cdot u^k]-g^{i\overline{l}}\cdot (g_{l%
\overline{k}}\cdot \,^QS^{jk})_{;j}\text{ .}
\end{array}
\end{equation}

By means of the covariant divergency of the energy-momentum tensors the
covariant Noether identities can be represented in index-free forms.

\section{Covariant Noether's identities and relations between their
structures}

The covariant Noether identities 
\[
\overline{F}_\alpha +\overline{\theta }_\alpha \,^\beta \,_{/\beta }\equiv 0%
\text{ , \thinspace \thinspace \thinspace \thinspace \thinspace \thinspace
\thinspace \thinspace \thinspace \thinspace \thinspace }\overline{F}_i+%
\overline{\theta }_i\,^j\,_{;j}\equiv 0\text{\thinspace ,\thinspace
\thinspace } 
\]
\[
\overline{\theta }_\alpha \,^\beta -\,_sT_\alpha \,^\beta \equiv \overline{Q}%
_\alpha \,^\beta \text{ ,\thinspace \thinspace \thinspace \thinspace
\thinspace \thinspace \thinspace \thinspace \thinspace \thinspace \thinspace
\thinspace \thinspace \thinspace \thinspace \thinspace \thinspace \thinspace 
}\overline{\theta }_i\,^j-\,_sT_i\,^j\equiv \overline{Q}_i\,^j\text{
,\thinspace \thinspace } 
\]

\noindent for the mixed tensor fields of second rank of the type 1. $\theta $%
, $_sT$ and $Q$ can be written in index-free form by the use of the
covariant divergency as 
\[
F+\delta \theta \equiv
0,\,\,\,\,\,\,\,\,\,\,\,\,\,\,\,\,\,\,\,\,\,\,\,\,\,\theta -\,_sT\equiv Q, 
\]
\begin{equation}  \label{X.6.-3}
\overline{g}(F)+\overline{g}(\delta \theta )\equiv
0,\,\,\,\,\,\,\,\,\,\,\,\,\,\,\,\,\,\,\,\,\,\,\,\,\,\,\,\,(\theta )\overline{%
g}-(\,_sT)\overline{g}\equiv (Q)\overline{g},
\end{equation}

\noindent where 
\begin{equation}  \label{X.6.-4}
F=\,_vF+\,_gF\text{ ,}
\end{equation}
\begin{equation}  \label{X.6.-8}
\begin{array}{c}
_vF=\,_{va}F+\,_vW \text{ , \thinspace \thinspace \thinspace \thinspace
\thinspace }_gF=\,_{ga}F+\,_gW\text{ ,\thinspace \thinspace \thinspace
\thinspace \thinspace \thinspace \thinspace }_aF=\;_{va}F+\,_{ga}F\text{
,\thinspace \thinspace } \\ 
W=\,_vW+\,_gW\text{ ,\thinspace }_vF=\,_v\overline{F}_\alpha \cdot e^\alpha
=\,_v\overline{F}_i\cdot dx^i\text{ ,\thinspace \thinspace \thinspace
\thinspace }_vW=\,_v\overline{W}_\alpha \cdot e^\alpha \text{ ,}
\end{array}
\end{equation}
\begin{equation}  \label{X.6.-9}
_{ga}F=\frac{\delta _gL}{\delta g_{\beta \gamma }}\cdot g_{\beta \gamma
/\alpha }\cdot e^\alpha \text{ ,\thinspace \thinspace \thinspace \thinspace
\thinspace \thinspace \thinspace \thinspace \thinspace \thinspace \thinspace 
}_gW=\,_g\overline{W}_\alpha \cdot e^\alpha \text{ .}
\end{equation}

From the second Noether identity ($\theta -\,_sT\equiv Q$) the relation
between the covariant divergencies of the energy-momentum tensors $\theta $, 
$_sT$ and $Q$ follows $\delta \theta \equiv \delta _sT+\delta Q$,\thinspace $%
\delta _sT\equiv \delta \theta -\delta Q$.

\begin{definition}
\textit{Local covariant conserved quantity }$G$\textit{\ of the type of an
energy-momentum tensor of the type 1. Mixed tensor field }$G$\textit{\ of
the type 1. with vanishing covariant divergency, i. e. }$\delta G=0$, $%
G_\alpha \,^\beta \,_{/\beta }=0$, $G_i\,^j\,_{;j}=0$.
\end{definition}

If a given energy-momentum tensor has to fulfil conditions for a local
covariant conserved quantity, then relations follow from the covariant
Noether identities (CNIs) between the covariant divergencies of the other
energy-momentum tensors and the covariant vector field $F$

\begin{center}
$
\begin{array}{cccc}
\text{No.} & \text{Condition for }\delta G & \text{Condition for }F & \text{%
Corollaries from CNIs} \\ 
1. & \delta \theta =0 & F=0 & \delta _sT=-\,\,\delta Q \\ 
2. & \delta _sT=0 & F\neq 0 & \delta \theta =\delta Q=-\,\,F \\ 
&  & F=0 & \delta \theta =\delta Q \\ 
3. & \delta Q=0 & F\neq 0 & \delta \theta =\delta _sT=-\,\,F \\ 
&  & F=0 & \delta \theta =\delta _sT=0
\end{array}
$
\end{center}

\textit{Special case}: 
\[
\frac{\delta _vL}{\delta V^A\,_B}=0,\,\,\,\,\,\,\,\,\,\,\,\,\,\,\,\,\,\,\,\,%
\,\,\,\,\,\,\,\,\,\,\,\frac{\delta _gL}{\delta g_{\alpha \beta }}=0\,\,\,%
\text{.} 
\]
\begin{equation}  \label{X.6.-11}
_{va}F=0\text{ ,\thinspace \thinspace \thinspace \thinspace \thinspace
\thinspace \thinspace \thinspace \thinspace \thinspace \thinspace \thinspace 
}_{ga}F=0\text{ ,\thinspace \thinspace \thinspace \thinspace \thinspace
\thinspace \thinspace \thinspace \thinspace \thinspace \thinspace }Q=0\text{
,\thinspace \thinspace \thinspace \thinspace \thinspace \thinspace
\thinspace \thinspace \thinspace \thinspace \thinspace \thinspace }_vF=\,_vW%
\text{ ,\thinspace \thinspace \thinspace \thinspace \thinspace \thinspace
\thinspace \thinspace \thinspace \thinspace \thinspace \thinspace \thinspace
\thinspace \thinspace \thinspace \thinspace \thinspace \thinspace }_gF=\,_gW%
\text{ ,}
\end{equation}
\begin{equation}  \label{X.6.-12}
_aF=\,_{va}F+\,_{ga}F=0\text{ ,}
\end{equation}
\begin{eqnarray}
F &=&W:W+\delta \theta =0\text{ ,\thinspace \thinspace \thinspace }\theta
=\,_sT\text{ ,\thinspace \thinspace \thinspace \thinspace }\delta \theta
=\delta _sT=-\,\,W\text{ .}  \label{X.6.-14} \\
\text{For\thinspace \thinspace \thinspace \thinspace }W &=&0:\delta \theta
=0,\delta _sT=0\text{ .}  \nonumber
\end{eqnarray}

The finding out the covariant Noether identities for a given Lagrangian
density $\mathbf{L}=\sqrt{-d_g}.L$ along with the energy-momentum tensors $%
\theta $, $_sT$ and $Q$ allow the construction of a rough scheme of the
structures of a Lagrangian theory over a differentiable manifold with
contravariant and covariant affine connections and a metric:%
%TCIMACRO{\TeXButton{newline}{\newline}}
%BeginExpansion
\newline%
%EndExpansion

\begin{center}
$
\begin{array}{ccccccccc}
& \leftarrow &  &  & \leftarrow \fbox{L}\rightarrow &  &  & \rightarrow & 
\\ 
\downarrow &  &  &  &  &  &  &  & \downarrow \\ 
&  &  &  &  &  &  &  &  \\ 
&  &  &  & \downarrow &  &  &  & \downarrow \\ 
&  &  &  & \fbox{$_s$T} &  &  &  & \fbox{$\delta $L/$\delta $K$^A$\thinspace 
$_B$} \\ 
\downarrow &  &  &  &  &  &  &  & \downarrow \\ 
\fbox{$\theta $} &  &  &  &  &  &  &  & \fbox{Q} \\ 
\downarrow &  &  &  &  &  &  &  & \downarrow \\ 
&  &  &  & \downarrow &  &  &  &  \\ 
\downarrow & \rightarrow &  & \rightarrow & \fbox{$\theta \,$-\thinspace $_s$%
T$\,\equiv \,$Q} & \leftarrow &  & \leftarrow & \downarrow \\ 
&  &  &  &  &  &  &  & \fbox{F} \\ 
\downarrow &  &  &  & \downarrow &  &  &  & \downarrow \\ 
& \rightarrow &  & \rightarrow & \fbox{F\thinspace +$\,\delta \theta \equiv
\,$0} & \leftarrow &  & \leftarrow & 
\end{array}
$

%TCIMACRO{\TeXButton{vspace{1mm}}{\vspace{1mm}}}
%BeginExpansion
\vspace{1mm}%
%EndExpansion
$\_\_\_\_\_\_\_\_\_\_\_\_\_\_\_\_\_\_\_\_\_\_\_\_\_\_\_\_\_\_\_\_\_\_\_\_\_%
\_\_\_\_\_\_\_\_\_\_\_\_\_\_$

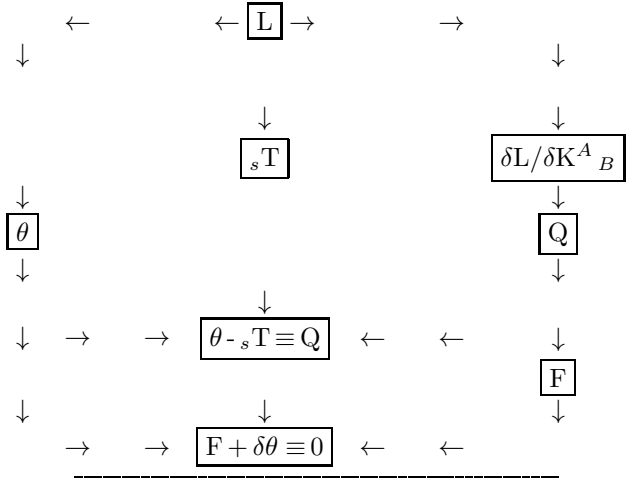
\textbf{Fig. 1}. \textit{Scheme of the main structure of a Lagrangian theory}
\end{center}

The symmetric energy-momentum tensor of Hilbert $_{gsh}T$ appears as a
construction related to the functional variation of the metric field
variables $g_{\alpha \beta }$ [as a part of the variables $K^A\,_B\sim
(V^A\,_B$, $g_C)$] and interpreted as a symmetric energy-momentum tensor of
a Lagrangian system. This tensor does not exist as a relevant element of the
scheme for obtaining Lagrangian structures by the method of Lagrangians with
covariant derivatives (MLCD). It takes in the scheme a separate place and
has different than the usual for the other elements relations.

\begin{center}
$
\begin{array}{cccccccccc}
& \leftarrow &  &  & \leftarrow \fbox{L}\rightarrow &  &  & \rightarrow & 
\rightarrow &  \\ 
\downarrow &  &  &  & \downarrow &  &  &  & \downarrow & \downarrow \\ 
&  &  &  &  &  &  &  &  &  \\ 
&  &  &  & \downarrow &  &  &  & \downarrow &  \\ 
&  &  &  & \fbox{$_s$T} &  &  &  & \fbox{$\delta $L/$\delta $K$^A$\thinspace 
$_B$} & \downarrow \\ 
\downarrow &  &  &  &  &  &  &  & \downarrow & \fbox{$_{gsh}$T} \\ 
\fbox{$\theta $} &  &  &  &  &  &  &  & \fbox{Q} &  \\ 
&  &  &  &  &  &  &  &  &  \\ 
\downarrow &  &  &  & \downarrow &  &  &  &  &  \\ 
& \rightarrow &  & \rightarrow & \fbox{$\theta \,$-\thinspace $_s$T$\,\equiv
\,$Q} & \leftarrow &  & \leftarrow & \downarrow &  \\ 
&  &  &  &  &  &  &  & \fbox{F} &  \\ 
\downarrow &  &  &  & \downarrow &  &  &  & \downarrow &  \\ 
& \rightarrow &  & \rightarrow & \fbox{F\thinspace +$\,\delta \theta \equiv
\,$0} & \leftarrow &  & \leftarrow &  & 
\end{array}
$

%TCIMACRO{\TeXButton{vspace{1mm}}{\vspace{1mm}}}
%BeginExpansion
\vspace{1mm}%
%EndExpansion
$\_\_\_\_\_\_\_\_\_\_\_\_\_\_\_\_\_\_\_\_\_\_\_\_\_\_\_\_\_\_\_\_\_\_\_\_\_%
\_\_\_\_\_\_\_\_\_\_\_\_$

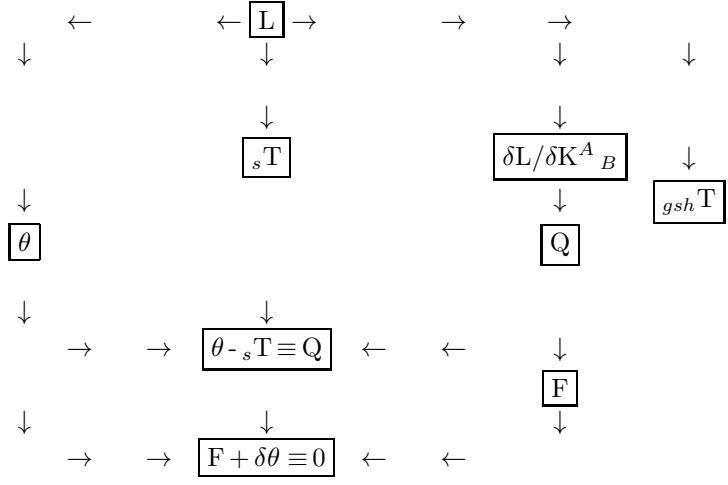
\textbf{Fig. 2}. \textit{The main structures of a Lagrangian theory }

\textit{and the energy-momentum tensor of Hilbert}
\end{center}

The field theories involve relations between the different structures of the
Lagrangian systems. For the most part of Lagrangian systems equations of the
type of the Euler-Lagrange equations have been imposed and the symmetric
energy-momentum tensor of Hilbert has been used. In Einstein's theory of
gravitation (ETG) the existing relations among the different structure's
elements are very peculiar. They require additional considerations.%
%TCIMACRO{\TeXButton{newline}{\newline}}
%BeginExpansion
\newline%
%EndExpansion

\begin{center}
$
\begin{array}{ccccccccc}
\fbox{L}\rightarrow &  & \rightarrow &  & \fbox{$\frac{\delta (\text{L}%
_g+\,\,\text{L})}{\delta g_{\alpha \beta }}=0$} &  & \leftarrow &  & 
\leftarrow \fbox{L$_g$} \\ 
\downarrow & \rightarrow & \frame{$\frac{\delta \text{L}}{\delta V^A\,_B}\,$%
\fbox{=\thinspace 0}} & \downarrow & \downarrow & \downarrow & \fbox{$\frac{%
\delta \text{L}_g}{\delta g_{\alpha \beta }}$} & \leftarrow & \downarrow \\ 
\fbox{$\theta $} &  &  & \fbox{$_{mvsh}$T} & \fbox{=} & \fbox{$-\,_{ggsh}$T}
&  &  & \fbox{$_g\theta $} \\ 
\fbox{$_s$T} &  &  &  &  &  &  &  & \fbox{$_{gs}$T}
\end{array}
$

%TCIMACRO{\TeXButton{vspace{1mm}}{\vspace{1mm}}}
%BeginExpansion
\vspace{1mm}%
%EndExpansion
$\_\_\_\_\_\_\_\_\_\_\_\_\_\_\_\_\_\_\_\_\_\_\_\_\_\_\_\_\_\_\_\_\_\_\_\_\_%
\_\_\_\_\_\_\_\_\_\_\_\_\_\_\_\_\_\_\_\_\_\_\_\_\_\_\_$

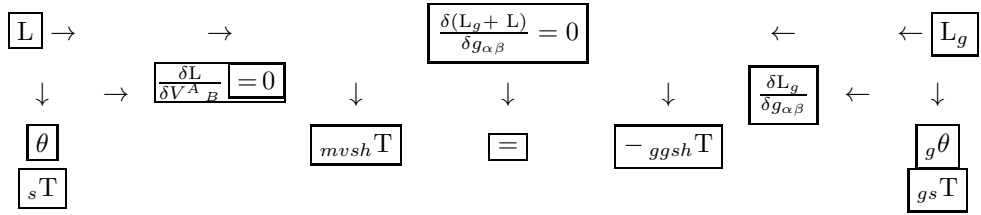
\textbf{Fig}.\textbf{\ 3.}\textit{\ Structure of the Einstein theory of
gravitation.}
\end{center}

In the Fig. 3. L is the Lagrangian invariant of the material distribution. L$%
_g$ is the Lagrangian invariant of the gravitational field.

There are other possible relations between the structures of two Lagrangian
densities than the relations between $\mathbf{L}_g$ and $\mathbf{L}_m$ in
ETG. For instance, the relations between the variational energy-momentum
tensors $_gQ$ and $_mQ$ of both Lagrangian densities and between the vector
fields $_gF$ and $_mF$.

\begin{center}
$
\begin{array}{ccccccc}
& \leftarrow \fbox{L}\rightarrow &  &  &  & \leftarrow \fbox{L$_0$}%
\rightarrow &  \\ 
\downarrow & \downarrow & \downarrow &  & \downarrow & \downarrow & 
\downarrow \\ 
&  & \fbox{$\delta $L/$\delta $K$^A$\thinspace $_B$} &  & \fbox{$\delta $L$%
_0 $/$\delta $G$^C$\thinspace $_D$} &  &  \\ 
& \downarrow &  &  &  & \downarrow &  \\ 
& \fbox{$_s$T} &  &  &  & \fbox{$_s$T$_0$} &  \\ 
\downarrow &  & \downarrow &  & \downarrow &  & \downarrow \\ 
\fbox{$\theta $} &  & \fbox{Q} & \fbox{=} & \fbox{Q$_0$} &  & \fbox{$\theta
_0$} \\ 
\downarrow & \downarrow &  &  &  & \downarrow & \downarrow \\ 
\rightarrow & \fbox{$\theta \,$-\thinspace $_s$T$\,\equiv \,$Q} &  &  &  & 
\fbox{$\theta _0\,$-\thinspace \thinspace $_s$T$_0\equiv \,$Q$_0$} & 
\leftarrow \\ 
\downarrow &  & \downarrow &  & \downarrow &  & \downarrow \\ 
&  & \fbox{F} & \fbox{=} & \fbox{F$_0$} &  &  \\ 
& \downarrow & \downarrow &  & \downarrow & \downarrow &  \\ 
\rightarrow & \fbox{F\thinspace +$\,\delta \theta \equiv \,$0} & \leftarrow
&  & \rightarrow & \fbox{F$_0\,$+$\,\delta \theta _0\equiv \,$0} & \leftarrow
\end{array}
$

%TCIMACRO{\TeXButton{vspace{1mm}}{\vspace{1mm}}}
%BeginExpansion
\vspace{1mm}%
%EndExpansion
$\_\_\_\_\_\_\_\_\_\_\_\_\_\_\_\_\_\_\_\_\_\_\_\_\_\_\_\_\_\_\_\_\_\_\_\_\_%
\_\_\_\_\_\_\_\_\_\_\_\_\_\_\_\_\_\_\_$

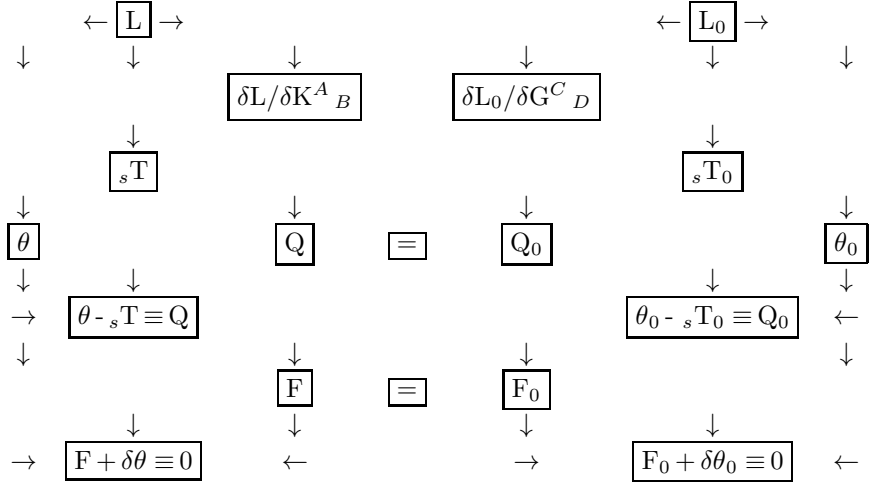
\textbf{Fig. 4.}\textit{\ Possible relations between two Lagrangian systems}
\end{center}

On the grounds of relations of this type a model for describing the
gravitational interaction in $V_4$-spaces is considered different from ETG 
\cite{Manoff-5}.

The covariant Noether identities can be used for a generalization of notions
of the continuum media mechanics related to the notions of force density,
density of the moment of the force (density of the inertial momentum) and
angular momentum density for Lagrangian systems described by models over
differentiable manifolds with affine connections and metrics.

\begin{center}
\textbf{Acknowledgments}
\end{center}

The author is grateful to Prof. St. Dimiev from the Institute for
Mathematics and Informatics of the Bulgarian Academy of Sciences, Prof. K.
Sekigawa from the Department of Mathematics of the Faculty of Science at the
Niigata University (Japan) for their support of the presented topics as well
as to Prof. A. N. Chernikov from the Bogoliubov Laboratory of Theoretical
Physics at the Joint Institute for Nuclear Research (Russia) and Prof. D. I.
Kazakov for the kind hospitality at the Joint Institute for Nuclear
Research. This work is supported in part by the National Science Foundation
of Bulgaria under Grant No. F-642.

%TCIMACRO{\TeXButton{small}{\small}}
%BeginExpansion
\small%
%EndExpansion

\end{document}